\documentclass[aps,preprint,showpacs,showkeys]{revtex4}
\usepackage{graphicx}
\usepackage{amsmath}
\usepackage{amsfonts}
\usepackage{amsthm}
\usepackage{mathrsfs}
\usepackage{amssymb}
\usepackage{graphicx}
\usepackage{epsfig}
\usepackage{hyperref}
\usepackage{amscd}
\begin{document}
\newcommand{\be}{\begin{equation}}
\newcommand{\ee}{\end{equation}}
\newcommand{\bea}{\begin{eqnarray}}
\newcommand{\eea}{\end{eqnarray}}

\title{Motion of test particles in the field of a naked singularity}

\author{K. Boshkayev$^{1,4}$, E. Gasper\'\i n$^2$, A. C. Guti\'errez-Pi\~neres$^{2,3}$, H. Quevedo$^{2,4}$ and S. Toktarbay$^1$}
\affiliation{
$^1$ Physical-Technical Faculty, Al-Farabi Kazakh National University, Al Farabi av. 71, 050040 Almaty, Kazakhstan\\
$^2$ Instituto de Ciencias Nucleares, 
Universidad Nacional Aut\'onoma de M\'exico, 
 AP 70543, M\'exico, DF 04510, Mexico\\ 
$^3$ Facultad de Ciencias B\'asicas,
Universidad Tecnol\'ogica de Bol\'ivar, Cartagena, Colombia\\ 
$^4$ Dipartimento di Fisica and ICRA, Universit\'a di Roma ``La Sapienza", Piazzale Aldo Moro 5, I-00185 Roma, Italy
}

\date{\today}

\begin{abstract}
We investigate the motion of test particles in the gravitational field of a static naked singularity generated by a mass distribution with quadrupole moment.
We use the quadrupole-metric ($q-$metric) which is the simplest generalization of the Schwarzschild metric with a quadrupole parameter. We study the influence 
of the quadrupole on the motion of massive test particles and photons and show that the behavior of the geodesics can drastically depend on the values 
of the quadrupole parameter. In particular, we prove explicitly that the perihelion distance depends on the value of the quadrupole. Moreover, we show that
 an accretion disk on the equatorial plane of the quadrupole source can be either continuous or discrete, depending on the value of the quadrupole. 
The inner radius of the disk can be used in certain cases to determine the value 
of the quadrupole parameter. The case of a discrete accretion is interpreted as due to the presence of repulsive gravity generated by the naked singularity. Radial geodesics are also investigated 
and compared with the Schwarzschild counterparts. 
\end{abstract}
\relax

\maketitle


\section{Introduction}
\label{sec:int}

The black hole uniqueness theorems state that in Einstein's general relativity theory the most general black hole solution
in empty space is described by the Kerr metric \cite{kerr63},
which represents the exterior gravitational field of a rotating 
mass $m$ with specific angular momentum  $a=J/m$. In Boyer-Lindquist coordinates $(t,r,\theta,\varphi)$, a true curvature 
singularity is determined by the equation $r^2 + a^2\cos^2\theta =0$ that corresponds to 
a ring singularity situated on the equatorial plane $\theta = \pi/2$. 
The ring singularity is isolated from the exterior 
spacetime by an event horizon situated on a sphere of radius 
$r_h = m + \sqrt{m^2-a^2}$. 

In the case that $a^2>m^2$, no event horizon exists 
and the ring singularity becomes naked. 
Different studies \cite{def78,cal79,rud98} show, however,
that in realistic situations, where astrophysical objects are mostly surrounded by accretion disks, a
Kerr naked singularity is an unstable configuration that 
rapidly decays into a Kerr black hole. Moreover, it now seems established that  in generic situations a 
gravitational collapse cannot lead to the formation of a
final configuration corresponding to a Kerr naked singularity.
These results seem to indicate that rotating Kerr naked singularities cannot be very common 
objects in nature. 

The above results seem to corroborate the validity of the cosmic censorship hypothesis \cite{penrose}
according to which a physically realistic gravitational collapse, which evolves from a regular initial state, 
can never lead to the formation of a naked singularity; that is, 
all singularities formed as the result of a realistic 
collapse should always be enclosed within an event horizon; hence, the singularities are always 
invisible to observers situated outside the horizon. Many attempts have been made 
to prove this conjecture with the same mathematical rigor used to show the inevitability of 
singularities in general relativity \cite{hawking}. No general proof has been formulated so far.
Instead, particular scenarios of gravitational collapses have been investigated some of which indeed 
corroborate the correctness of the conjecture. 

Nevertheless, other  studies \cite{naked} 
indicate that under certain circumstances naked singularities can appear as the result of 
a realistic gravitational collapse. Indeed, it turns out that in the case of the collapse of  an inhomogeneous matter distribution,  
there exists a critical degree of inhomogeneity below which black holes 
form. Naked singularities appear if the degree of inhomogeneity is higher than the critical value.
The speed of the collapse and the shape of the collapsing object are also factors that play an 
important role in the determination of the final configuration. It turns out that naked singularities form 
more frequently if the collapse occurs very rapidly and if the object is not  spherically
symmetric.

In view of this situation it seems reasonable to investigate the effects of naked singularities 
on the surrounding spacetime. This is the main aim of the present work. We will study a 
naked singularity without black hole counterpart. In fact, we investigate 
the quadrupole metric ($q-$metric) which is the  simplest generalization of the Schwarzschild metric containing a naked singularity. 
We will see that, starting from the Schwarzschild metric,
 the Zipoy--Voorhees \cite{zip66,voor70} transformation can be used 
to generate a static axisymmetric spacetime which describes the field of a mass 
with a particular quadrupole moment. For any values of the quadrupole, the spacetime
is characterized by the presence of naked singularities situated at a finite distance
from the origin of coordinates. In this work, we are interested in analyzing the spacetime outside the
outer naked singularity.

We perform an analysis of the motion of test particles (massive and non-massive) outside the naked singularity, 
comparing in all the cases our results with the corresponding situation in the Schwarzschild spacetime in order to 
establish the exact influence of the quadrupole on the parameters of the trajectories. 

This paper is organized as follows. In Sec. \ref{sec:qme}, we present the $q-$metric and investigate its main 
physical properties. It is shown that it can be used to describe the exterior gravitational field of a mass
distribution with quadrupole moment. Moreover, we prove that the corresponding spacetime is characterized by 
the presence of naked singularities. In Sec. \ref{sec:geo}, we derive the geodesic equations in general and
study numerically the conditions under which motion is allowed. Sec. \ref{sec:eqg} is devoted to the study 
of circular and non-circular equatorial geodesics. We show that the geometric configuration of an accretion disk 
located around a naked singularity depends explicitly on the value of the quadrupole in a such a way that it is 
always possible to distinguish between a Schwarzschild black hole and a naked singularity. This effect was found
previously in the field of naked singularities with electric charge \cite{pqr11a,pqr11b}. 
Furthermore, in Sec. \ref{sec:rad}, 
we study the influence of the quadrupole on the behavior of radial timelike and null geodesics. Finally, in Sec. \ref{sec:con}, 
we discuss our results and comment on possible future works.


\section{The $q-$metric}
\label{sec:qme}

In 1917 \cite{solutions}, it was shown that the most general static axisymmetric asymptotically flat solution of Einstein's vacuum equations is 
represented by the Weyl class.  
In terms of multipole moments, the simplest static 
solution contained in the Weyl class is the Schwarzschild metric which is the only one that possesses a mass monopole moment only. From a physical 
point of view, the next interesting solution should describe the exterior field of a mass with quadrupole moment. In this case, it is possible to find a large number of exact solutions with the same quadrupole (see \cite{quev10} and the references cited therein) that differ only in the set of higher multipoles. A common characteristic of the solutions with quadrupole is that their explicit form is rather cumbersome, making them difficult to be handled analytically \cite{quev90}.
An alternative exact solution was presented by one of us in \cite{quev11} by applying on the Schwarzschild metric a Zipoy-Voorhees transformation with 
parameter $\delta = 1 + q$, where $q$ represents the quadrupole parameter. 
In spherical coordinates, the resulting metric can be written in a compact and simple form as  
\bea
ds^2 = &&  \left(1-\frac{2m}{r}\right) ^{1+q} dt^2\nonumber  \\
&& - \left(1-\frac{2m}{r}\right)  ^{-q}\left[ \left(1+\frac{m^2\sin^2\theta}{r^2-2mr }\right)^{-q(2+q)} \left(\frac{dr^2}{1-\frac{2m}{r} }+ r^2d\theta^2\right) + r^2 \sin^2\theta d\varphi^2\right] \ .
\label{zv}
\eea

In the literature, this solution is known as the  $\delta-$metric or as the $\gamma-$metric for notational reasons \cite{mala04}. We propose to use the term 
quadrupole metric ($q-$metric) to emphasize the role of the parameter $q$ which determines the quadrupole moment, as we will see below.  

The $q-$metric is an axially symmetric exact vacuum solution that reduces to the spherically symmetric Schwarzschild metric only for $q\rightarrow 0$. 
It is asymptotically flat for any finite values of the parameters $m$ and $q$. Moreover, in the limiting case $m\rightarrow 0$
it can be shown that, independently of the value of $q$, there exists a coordinate transformation that 
transforms the resulting metric into the Minkowski solution. This last property is important from a physical point of view 
because it means that the parameter $q$ is related to a genuine mass distribution. 

An important quantity that characterizes the physical properties of any exact solution is the 
Arnowitt-Deser-Misner (ADM) mass. If one tries to use the common formula \cite{wald,tod11} for calculating the ADM mass, one would obtain incorrect results because these formula are adapted to a particular coordinate system. Therefore, we perform here a detailed calculation of the ADM mass, using the original approach. 
All the details are given in the 
Appendix. In the case of the $q-$metric (\ref{zv}), the final result is simply 
\be
M_{ADM} = m(1+q) \ .
\label{adm}
\ee
It follows that if we take the parameter $m$ as positive, the parameter $q$ must satisfy the condition $q>-1$ in order for the ADM mass to be positive. 
Nevertheless, one can choose a negative $m$ so that for $q<-1$ the mass is still positive. For the sake of simplicity we assume from now on that $m$ is positive.   

We also calculate the multipole moments of the $q-$metric by using the invariant definition proposed by Geroch \cite{ger}. The 
lowest mass multipole moments $M_n$, $n=0,1,\ldots $ are given by
\be 
M_0= (1+q)m\ , \quad M_2 = -\frac{m^3}{3}q(1+q)(2+q)\ ,
\ee
whereas higher moments are proportional to $mq$ and can be 
completely rewritten in terms of $M_0$ and $M_2$. This means that the arbitrary parameters $m$ and $q$ determine the mass and quadrupole 
which are the only independent multipole moments of the solution. In the limiting case $q=0$ only the monopole $M_0=m$ 
survives, as in the Schwarzschild spacetime. In the limit $m\rightarrow 0$, with $q\neq 0$, and $q\rightarrow -1$, with $m\neq 0$, 
 all multipoles vanish identically, implying that 
no mass distribution is present and the spacetime must be flat. 
Furthermore, notice that all odd multipole moments are zero because the solution possesses an additional 
reflection symmetry with respect to the equatorial plane $\theta=\pi/2$. 

We conclude that the above metric describes the exterior gravitational 
field of a static deformed mass. The deformation is described by the quadrupole moment $M_2$ which is positive for a prolate mass 
distribution and negative for an oblate one. Notice that the condition $q>-1$ must be satisfied in order to avoid the appearance of
a negative total mass $M_0$. Therefore, in the interval $q\in (-1,0)$ the $q-$metric describes a prolate mass distribution and in the interval
$(0,\infty)$ an oblate one.

\begin{figure}
\includegraphics[scale=0.5]{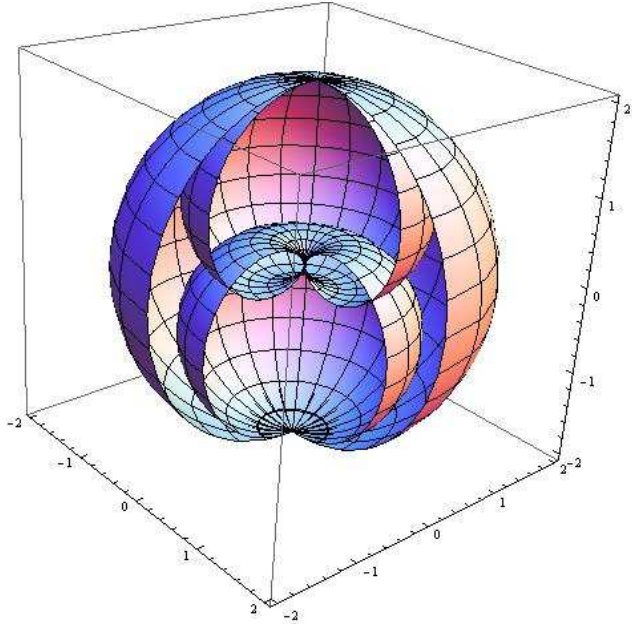}%
\caption{Curvature singularities of the $q-$metric. The outer singularity at $r=2m$ and the singularity at the origin $r=0$ exist for any value of the parameter $q$.
The singularity which depends on the angle $\theta$ exists only for certain values of the quadrupole (see text).}%
\label{fig1}%
\end{figure}

To investigate the structure of possible curvature singularities, we consider the Kretschmann scalar 
$K = R_{\mu\nu\lambda\tau}R^{\mu\nu\lambda\tau}$. A straightforward computation leads to 
\be
K   = \frac{16 m^2(1+q)^2}{r^{4(2+2q+q^2)}}\frac{ (r^2-2mr+m^2\sin^2\theta)^{2(2q+q^2)-1}}{(1-2m/r)^{2(q^2+q+1)}}L(r,\theta)\ ,
\label{kre}
\ee
with 
\bea
L(r,\theta)= & & 3(r-2m-qm)^2(r^2-2mr+m^2\sin^2\theta) \nonumber\\
& & + q(2+q)m^2\sin^2\theta[ q(2+q)m^2 + 3(r-m)(r-2m-qm)] \ .
\eea
In the limiting case $q=0$, we obtain the Schwarzschild value $K= {48 m^2}/{r^6}$ with the only singularity 
situated at the origin of coordinates $r\rightarrow 0$. 
In general, we can see that the singularity at 
the origin, which occurs at $r^{4(2+2q+q^2)} =0$,  is present for all real values of $q$. 
Moreover, an additional singularity appears at the radius $r=2m$ 
which, according to the form of the metric (\ref{zv}), is also a horizon in the sense that the norm of the timelike Killing 
vector vanishes at that radius. Outside the hypersurface $r=2m$ no additional horizon exists, indicating 
that the singularities situated at the origin and at $r=2m$ are naked. In addition, a curvature singularity occurs at the surface
determined by the equation
\be
r^2-2mr+m^2\sin^2\theta = 0
\ee
under the condition that the value of the quadrupole parameter is within the interval
 $q\in (-1,-1+\sqrt{3/2}]\backslash \{0\}$. We conclude that the main consequence of the existence of a quadrupole determined by the parameter $q$ is that
the Schwarzschild horizon becomes a naked singularity. The geometric structure of the curvature singularities of the $q-$metric is illustrated in Fig. \ref{fig1}.

Several physical properties of the $q-$metric have been investigated in the literature 
(see, for instance, \cite{par85,zv1,zv2,zv3,zv4,zv5,mala04,quev11,cpmj12} and references therein). In this work, we are interested in continuing the analysis of the
corresponding gravitational field as observed by exterior test particles.

	
\section{Geodesic motion}
\label{sec:geo}

Consider the trajectory $x^{\alpha}(\tau)$ of a test particle with 4-velocity   $u^{\alpha} =dx^{\alpha}/d\tau = {\dot x}^{\alpha}$.  
The  moment  $p^\alpha = \mu {\dot x}^{\alpha} $ of  the particle can be  normalized so  that 
   \begin{eqnarray}
    g_{\alpha\beta} \dot x^{\alpha} \dot x^{\beta} = \epsilon,\label{eq:normomentum} 
   \end{eqnarray}
where $\epsilon =0, 1, -1$ for null,  timelike, and  spacelike curves, respectively \cite{pqr11a}.  
For  the  $q-$metric we  obtain from (\ref{eq:normomentum}) that
   \begin{eqnarray}
   \left( 1 + \frac{m^2 \sin^2{\theta}}{r^2 - 2 m r}  \right)^{-q(2 + q)} {\dot r}^2 = {\tilde E}^2 - \Phi^2,\label{eq:genequat}
   \end{eqnarray}
where
   \begin{eqnarray}
   \Phi^2= \left( 1 -\frac{2m}{r} \right)^{1 + q} \left[   r^2 \left( 1 -\frac{2m}{r} \right)^{- q} \left( 1 + \frac{m^2 \sin^2\theta}{r^2 - 2m r}\right)^{-q(2     +   q)} {\dot\theta}^2 + \frac{{\tilde l}^2}{r^2 \sin^2 \theta} \left( 1 - \frac{2m}{r}\right)^q  + \epsilon
\right],
   \end{eqnarray}
and  we  have used the expression for  the energy $E=\mu \tilde E$ and the angular  moment $l=\mu \tilde l$ of  the test particle which are constants of  motion 
    \begin{eqnarray}
    E &=& g_{\alpha\beta}\xi^{\alpha}_{t}p^{\beta}= \left(1 - \frac{2m}{r}\right)^{1 + q} \mu {\dot t},\\
    l &=& - g_{\alpha\beta}\xi^{\alpha}_{\varphi}p^{\beta} = \left(1 - \frac{2m}{r}\right)^{- q} r^2 \sin^2 \theta \mu {\dot \varphi},
     \end{eqnarray}
associated with the  Killing vector  fields $\xi_t= \partial_t$ and $\xi_{\varphi}= \partial_{\varphi}$, respectively. For the sake of simplicity we set $\mu=1$ so that $\tilde E = E$ and $\tilde l = l$. 
     
In addition, the equation for the acceleration along the polar angle can be expressed as
\bea
\label{ddtheta}
         \ddot{\theta} &= 
                     -\left(\frac{ q(2 + q)m^2\sin{\theta}\cos{\theta}}{r^2 -2mr + m^2\sin^2{\theta}}\right)
                     \left( \frac{ \dot{r}^2 }{r^2 - 2mr} - \dot{\theta}^2 \right)
                    + \bigg( \frac{r^2 -2mr + m^2\sin^2{\theta}}{r^2 -2mr} \bigg)^{q(2+q)} 
                     \sin{\theta}\cos{\theta} \; \dot{\varphi}^2
                     \nonumber\\
                   & -   \frac{r^3 - (4+q)m r^2 +  \left[ 2(2 + q) + (1+q)^2\sin^2\theta \right]m^2 r
                     - (1+q)(2+ q)m^3 \sin^2\theta}{(r^2 -2mr)(r^2 -2mr + m^2\sin^2\theta)} 
                     \; \dot{r}\;\dot{\theta}.
\eea
From this equation it follows that the acceleration $\ddot \theta$ depends on the polar angle. Consider for instance the motion around the central object 
with initial values $\dot r = \dot \theta=0$ and $\dot\varphi \neq 0$. Then, Eq.(\ref{ddtheta}) reduces to
\be
\left( 1 + \frac{m^2\sin^2\theta}{r^2 - 2 m r} \right)^{-q(2+q)} \ddot \theta - \sin\theta\cos\theta \dot \varphi ^2  = 0 \ ,
\ee
It then follows that the only places at which the acceleration vanishes are $\theta = 0$ and $\theta= \pi/2$, showing that the polar and equatorial planes are geodesic planes at which the investigation of the test particle motion can be considerably simplified due to the symmetry of the configuration. For all the remaining values of 
$\theta$, there is always a non-zero value of the acceleration that induces a velocity in the direction of $\theta$ towards the equatorial plane or the poles. 

\begin{figure}%
\includegraphics[scale=0.4]{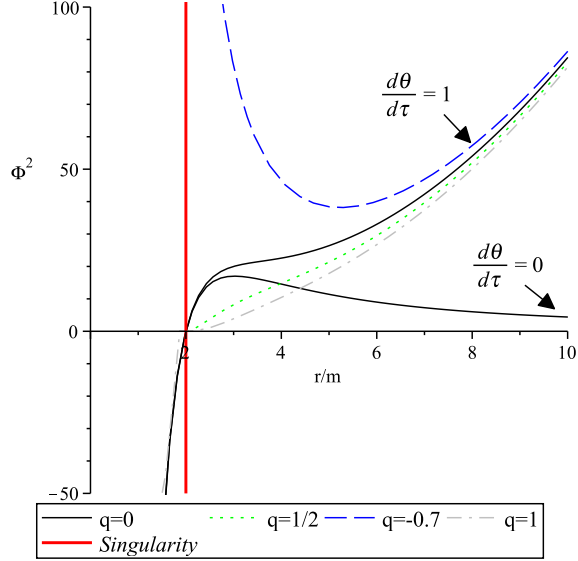}%
\caption{Behavior of the ``effective potential" $\Phi^2$ in terms of the radial distance $r$ for the particular velocity $\dot\theta=1$ with $\theta=\pi/4$, $l=15$ 
and different values of $q$. The effective potential for the Schwarzschild solution $q=0$ on a fixed plane with $\dot\theta=0$ is also depicted for comparison. }%
\label{fig2}%
\end{figure}

Equation (\ref{eq:genequat}) can be considered as describing the motion along the radial coordinate in terms of the ``effective potential" $\Phi^2$. However, the velocity $\dot\theta$ enters the expression for $\Phi^2$ explicitly so that strictly speaking it is not a potential. Nevertheless, its dependence on the coordinates can shed some light on the  behavior of the geodesics for particular values of $\dot\theta$. In Fig. \ref{fig2}, we plot the behavior of $\Phi^2$ for a particular initial velocity $\dot\theta =1$ and different values of the $q$ parameter. We see that at the singularity $r=2m$ the behavior of the potential drastically depends on the value of $q$. Indeed, for positive values of $q$ the potential either tends to infinity near the singularity or crosses it and becomes divergent inside the singularity. Instead, if $q$ is negative, the potential tends to infinity as the  singularity is approached from outside. In the case of vanishing quadrupole, the singularity turns into a horizon and the potential crosses it until it becomes infinity near the central singularity located at the origin of coordinates.

\begin{figure}%
\includegraphics[scale=0.4]{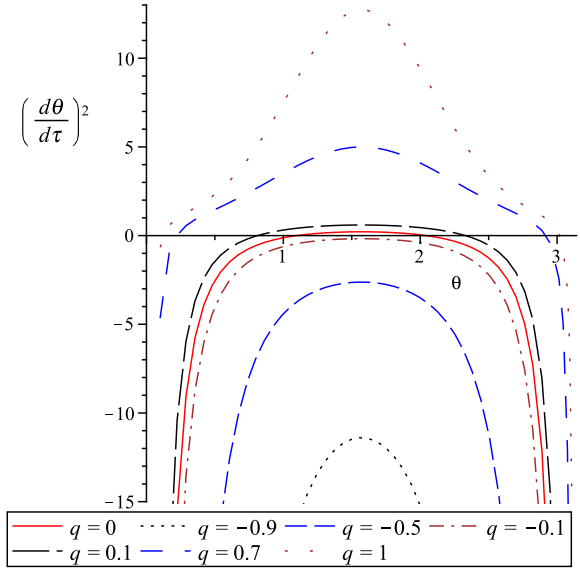}%
\caption{Angular velocity $\dot\theta$ in terms of the angle $\theta\in (0,\pi)$ for different quadrupoles. Here, we consider timelike geodesics ($\epsilon=1)$ 
with $m=1$, $ l= 5$. For concreteness, the radius of the orbit has been set to $r=2.35$.  }%
\label{fig2a}%
\end{figure}

An interesting particular case of the geodesic motion is that of circular orbits and their stability. The geodesic equation in this case is given as in 
Eq.(\ref{eq:genequat}) with $\dot r =0$, i.e. $\Phi^2 =  E ^2$, whose solutions strongly depend on the value of the velocity $\dot\theta$. In fact, 
the condition for circular orbits can be expressed as 
\be
{\dot \theta}^2= (r^2 - 2mr)^{-1}\left( 1 + \frac{m^2\sin^2\theta}{r^2 - 2mr}\right)^{q(2+q)}
\left[ { E}^2 - \frac{{ l}^2}{r^2\sin^2\theta} \left( 1 - \frac{2m}{r}\right)^{1 + 2q} - \epsilon \left( 1 - \frac{2m}{r}\right)^{1+q}     \right].
\ee
We perform a numerical investigation of this condition for a particular radius and different values of the quadrupole parameter in Fig. \ref{fig2a}. 
If we consider, for instance, the case of a circular orbit with radius $r=2.35m$, we can see that such a motion is not possible for arbitrary values of 
$q$ and $\theta$.  In fact, for most negative values of $q$ no circular orbit with this radius is allowed, because $\dot\theta^2$ is negative for all values of $\theta$. Instead, for all positive values of $q$ it is always possible to find an interval of $\theta$ within which it is possible to have circular orbits. 
For other values of the orbit radius similar results can be found.

The study of the stability of circular orbits is important to establish the possibility of having accretion disks around the central gravitational source. In particular, the radius of the last stable circular orbit determines the inner radius of the accretion disk, and corresponds to an inflection point of the potential, i.e., the point where the conditions $\partial \Phi^2/\partial r =0$  and $\partial^2 \Phi^2/\partial r^2 =0$ are satisfied. 
We use these two conditions to find the explicit values of $l^2$ and $\dot\theta^2$ for the last stable circular orbit and obtain 
\be
l_{lsco}^2=  \frac{\epsilon m (1+q) \sin^2{\theta}\;  r^2 }{r -(3+2q)m} 
               \left(1 - \frac{2m}{r}\right)^{-q} \left[1 - \frac{(r-m)\, G }
                              {r (r-2m) \left[r - (3 + 2q)m\right] H} \right],
\ee

and
\begin{align}
   \dot{\theta}_{lsco}^{2} = - \frac{ \epsilon m (1+q) (1-\frac{2m}{r})^{q-1} 
                            (r^2 -2mr + m^2\sin^2\theta)^{(1+ 2q + q^2)} G
                            }
                           {r^4 [r - (3+2q) m](r^2-2mr)^{q(2+q)} 
                           \left[r^2 - 2mr + m^2(1+ q)^2 \sin^2{\theta} \right] H},
   \end{align}
   
where
\be
   G\equiv r^2 - 2m(4+ 3q)r +  2m^2(2+ q)(3+ 2q) \ ,
\ee
   \bea
    H &\equiv 1+ \frac{r-m}{(r^2- 2mr) [r - (3+2q)m)]}  \bigg[3r^2 - 6m(3+2q)r + 4m^2(2+q)(3+2q) \nonumber\\
      & + \frac{2q(2+q)(1+q)^2m^4\sin^4{\theta}(r-m)[r -(3+2q)m]}
       { (r^2 - 2mr + m^2 \sin^2{\theta})[r^2 - 2mr + m^2(1+ q)^2 \sin^2{\theta}]}
       \bigg].
   \eea

\begin{figure}%
\includegraphics[scale=0.3]{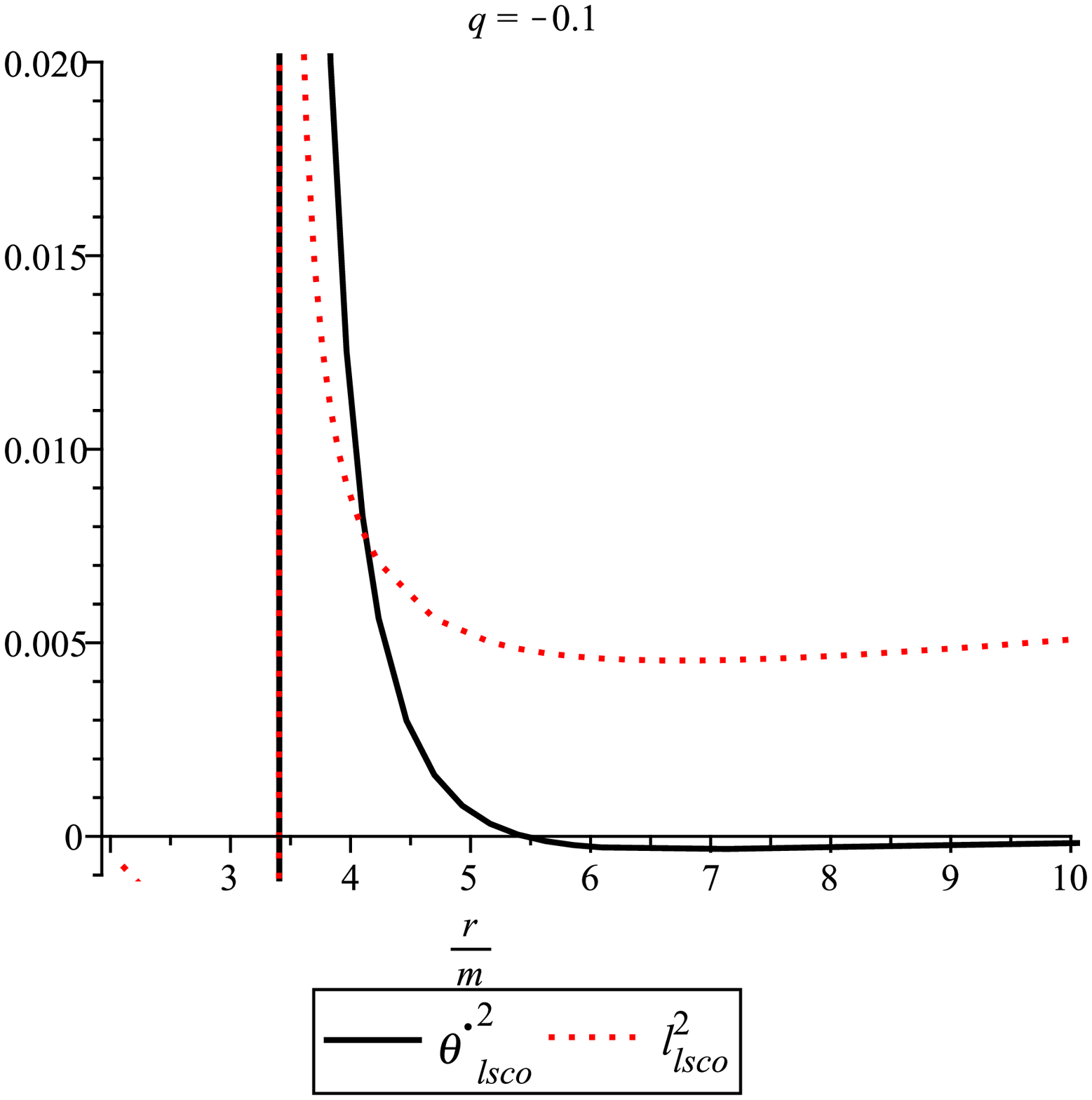}%
\includegraphics[scale=0.3]{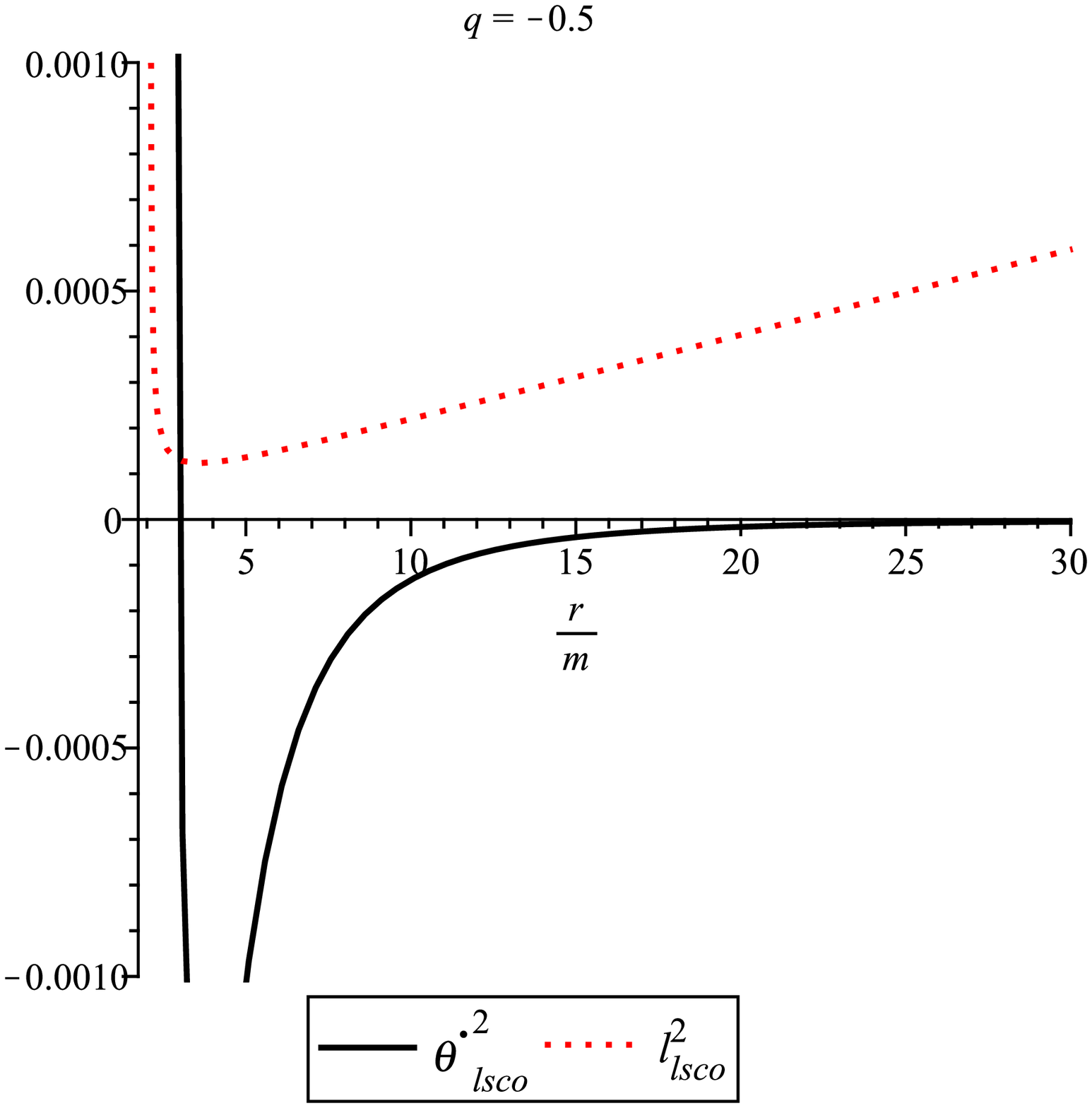}%
\\
\includegraphics[scale=0.3]{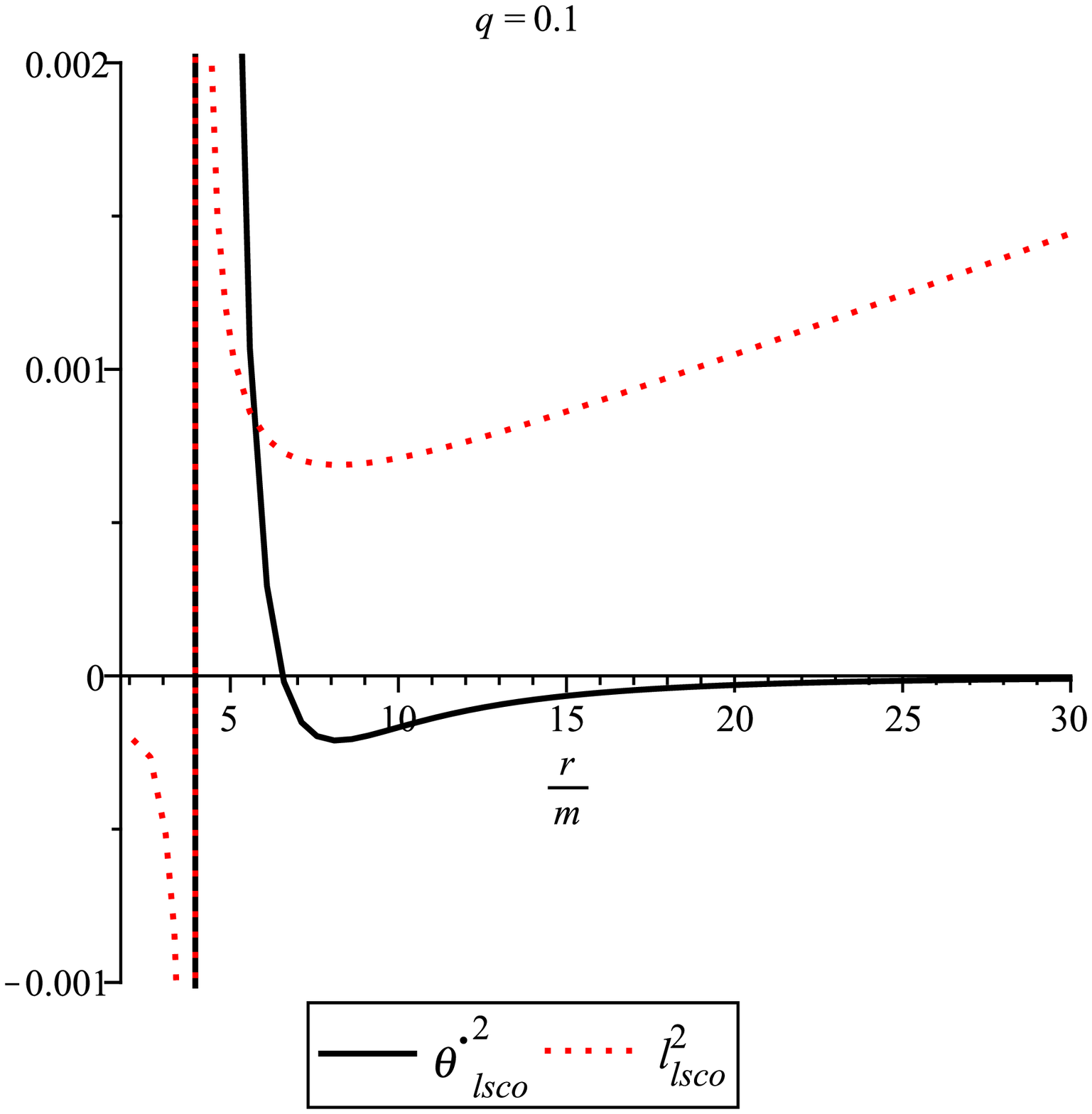}%
\includegraphics[scale=0.3]{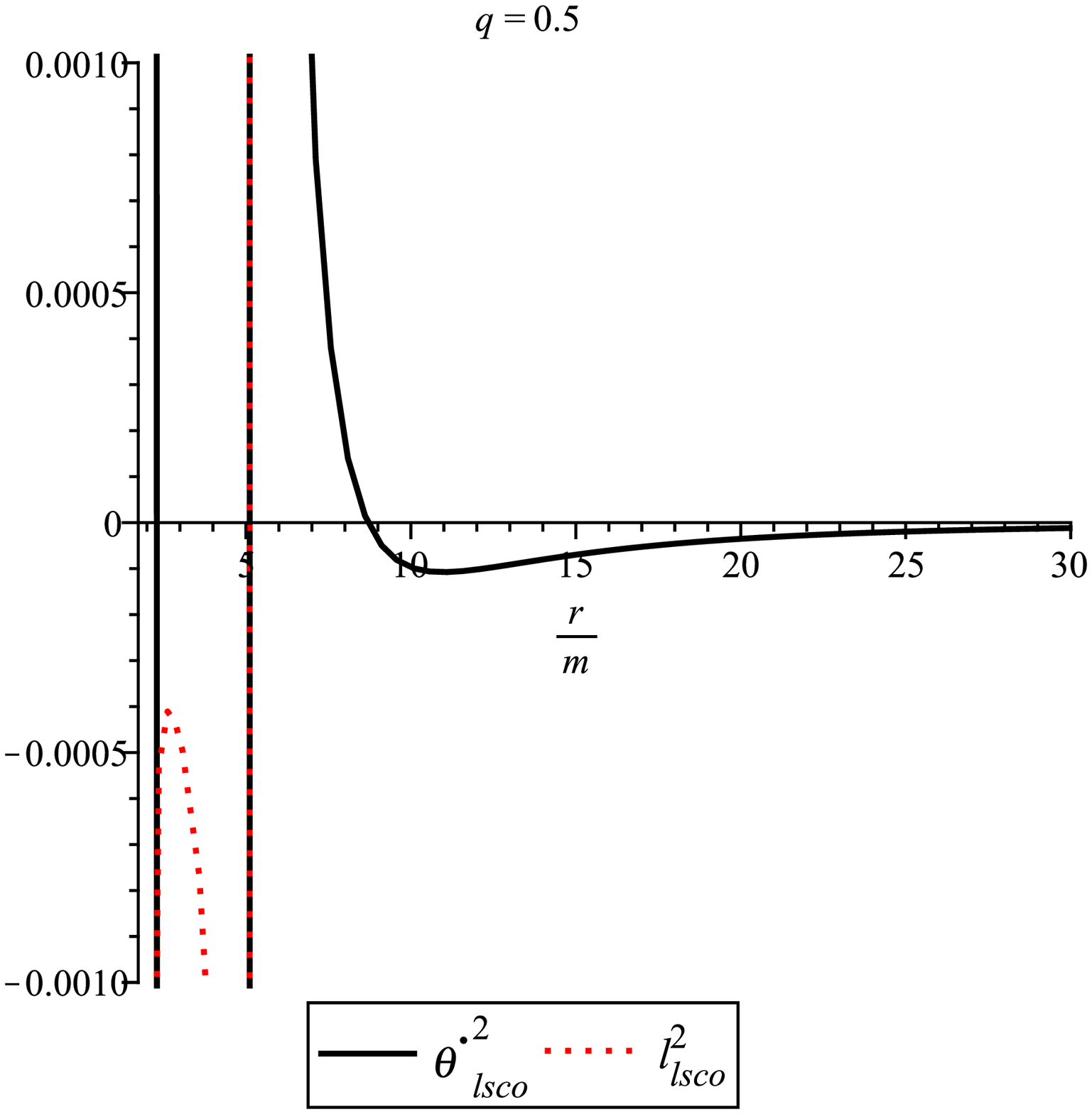}%
\caption{Angular momentum and polar angle velocity of the last stable circular timelike orbit as a function of the radial coordinate for different values of the quadrupole parameter $q$. Here we set $\theta=\pi/4$. The value of $l^2_{lsco}$ has been rescaled in all the graphs for comparison with the graph 
of $\dot\theta^2_{lsco}$.}  
\label{fig2b}%
\end{figure}

In the limiting case of the Schwarzschild spacetime ($q=0$), due to the spherical symmetry we can set $\dot\theta_{lsco}=0$ and so we obtain the value 
$r^{Sch}_{lsco}=6m$, as expected. The value of $l^2_{lsco}$ determines the angular momentum of the test particle on  the last stable circular orbit and 
$\dot\theta^2_{lsco}$ the velocity of the orbit with respect to the polar angle. In Fig. \ref{fig2b}, we analyze numerically the above equations for different values 
of the quadrupole. The condition for the existence of a last stable circular orbits is that both $l_{lsco}^2$ and $\dot\theta^2_{lsco}$ be positive and finite. 
From Fig. \ref{fig2b} we see that for $q=0.5$ and 
$q=-0.5$ these conditions are not satisfied.  For $q=0.1$ and $q=-0.1$, however, there exists an interval of $r$ in which both $l_{lsco}^2$ and $\dot\theta^2_{lsco}$ 
are positive.
The first thing one can notice is that this interval is always inside the Schwarzschild radius $r^{Sch}_{lsco}=6m$. This means that the quadrupole moment diminishes the 
value of the radius of the last stable circular orbit. For a particular positive value of $\dot\theta^2_{lsco}$ one can find from the graph the corresponding radius 
$r_{lsco}$. If this value happens to correspond to a positive value of $l_{lsco}^2$, then there exists a stable circular orbit with that radius and that angular momentum.

The motion of test particles for arbitrary values of $\dot\theta$ and $\theta$ can be investigated by performing a numerical integration of the geodesic equations
(\ref{eq:genequat})--(\ref{ddtheta}). Several cases must be considered, depending on the initial velocity $\dot\theta$ and the initial plane $\theta$. This requires a detailed numerical analysis in which several aspects must be considered like the stability of the trajectories and the appearance of chaotic motion. We will present these results elsewhere. In this work, we will focus on the analytic investigation of the most  important families of geodesics of the $q-$ metric.


\section{Equatorial geodesics}
\label{sec:eqg}

\begin{figure}%
\includegraphics[scale=0.4]{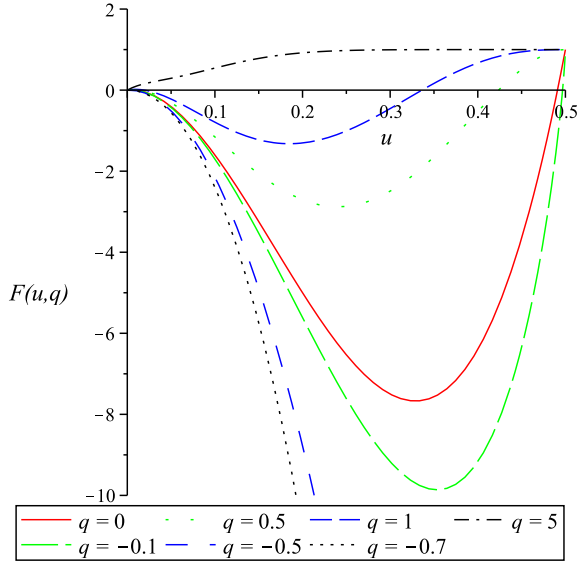}%
\caption{Behavior of the function $F(u,q)$ for  particular values of the parameters $m=1$, $  E=1$, and $  l =15$, and $\epsilon=1$.  }%
\label{fig3}%
\end{figure}

\begin{figure}%
\includegraphics[scale=0.4]{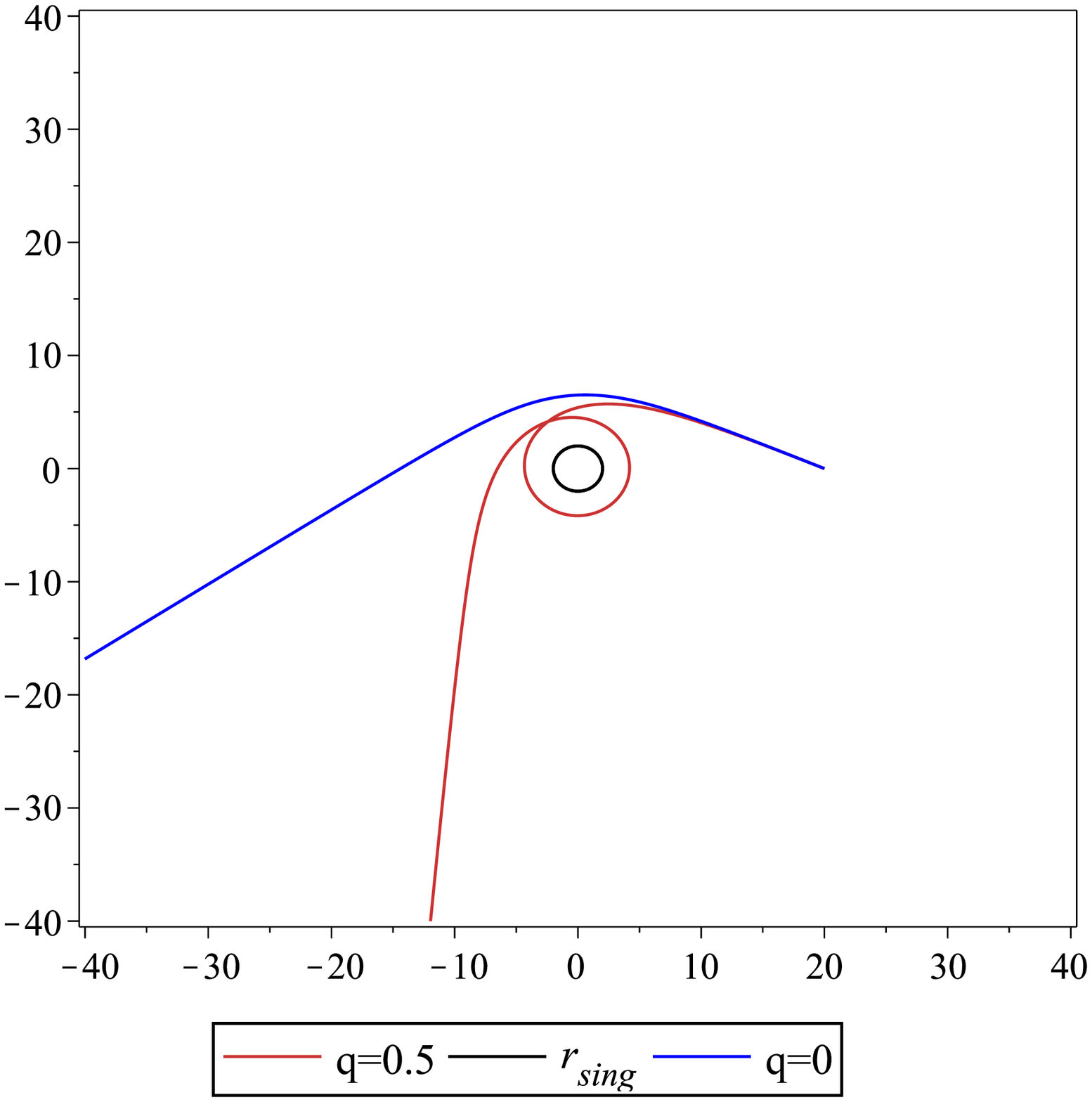}%
\quad 
\includegraphics[scale=0.4]{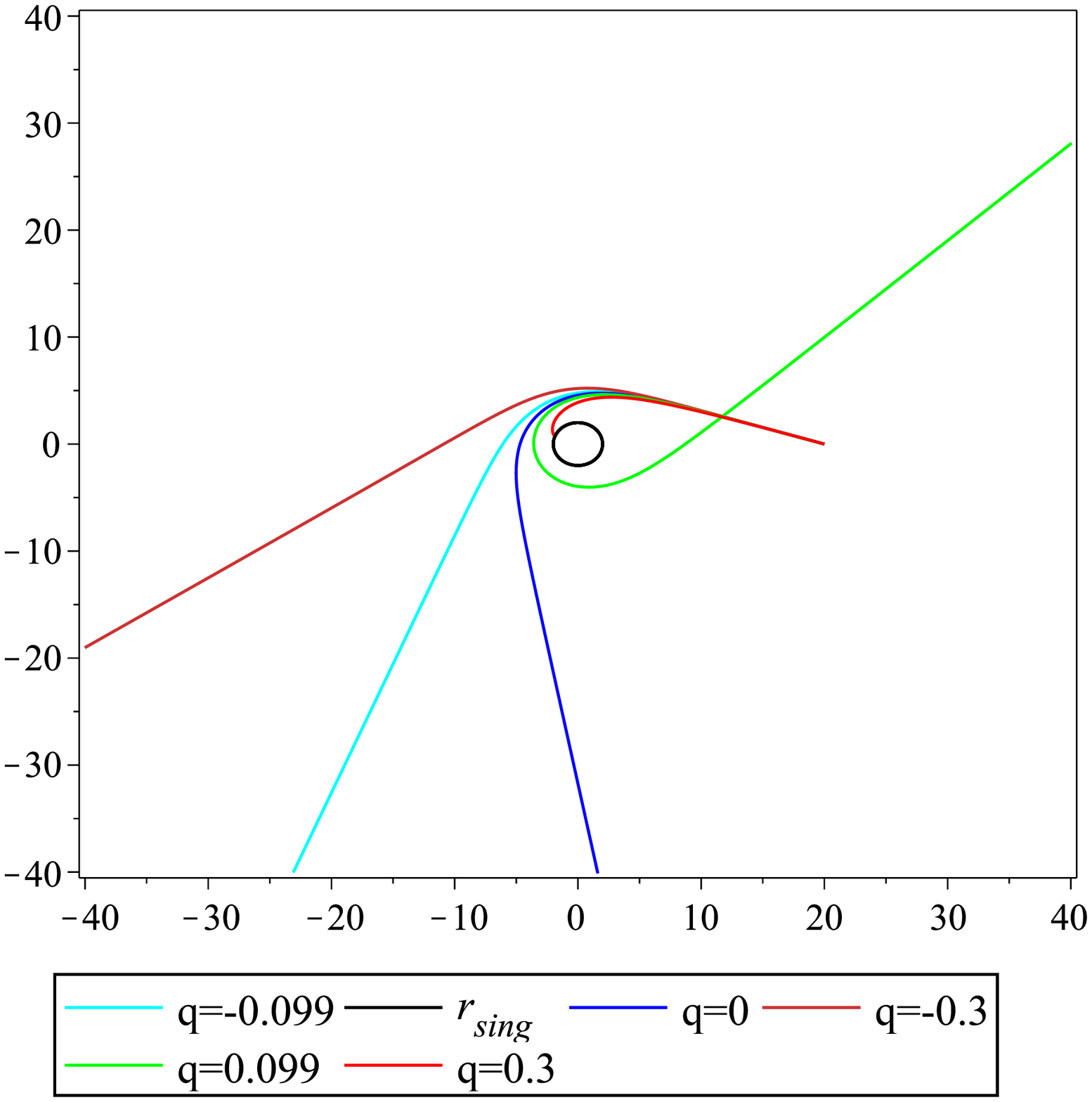}%
\caption{Influence of the quadrupole on unbounded Schwarzschild orbits with non-vanishing initial radial velocity ($\dot r(0)\neq 0)$.
The initial point 
$\varphi(0)=0$ and  $r(0)=20$ is the same for both plots. Initial velocities: $ \dot{r}(0)=-2$ and $\dot{\varphi}(0)=0.04305$ (left plot); 
 $\dot{r}(0)=-2.5$ and  $ \dot{\varphi}(0)=0.039$ (right plot).}
\label{fig4a}%
\end{figure}

\begin{figure}%
\includegraphics[scale=0.25]{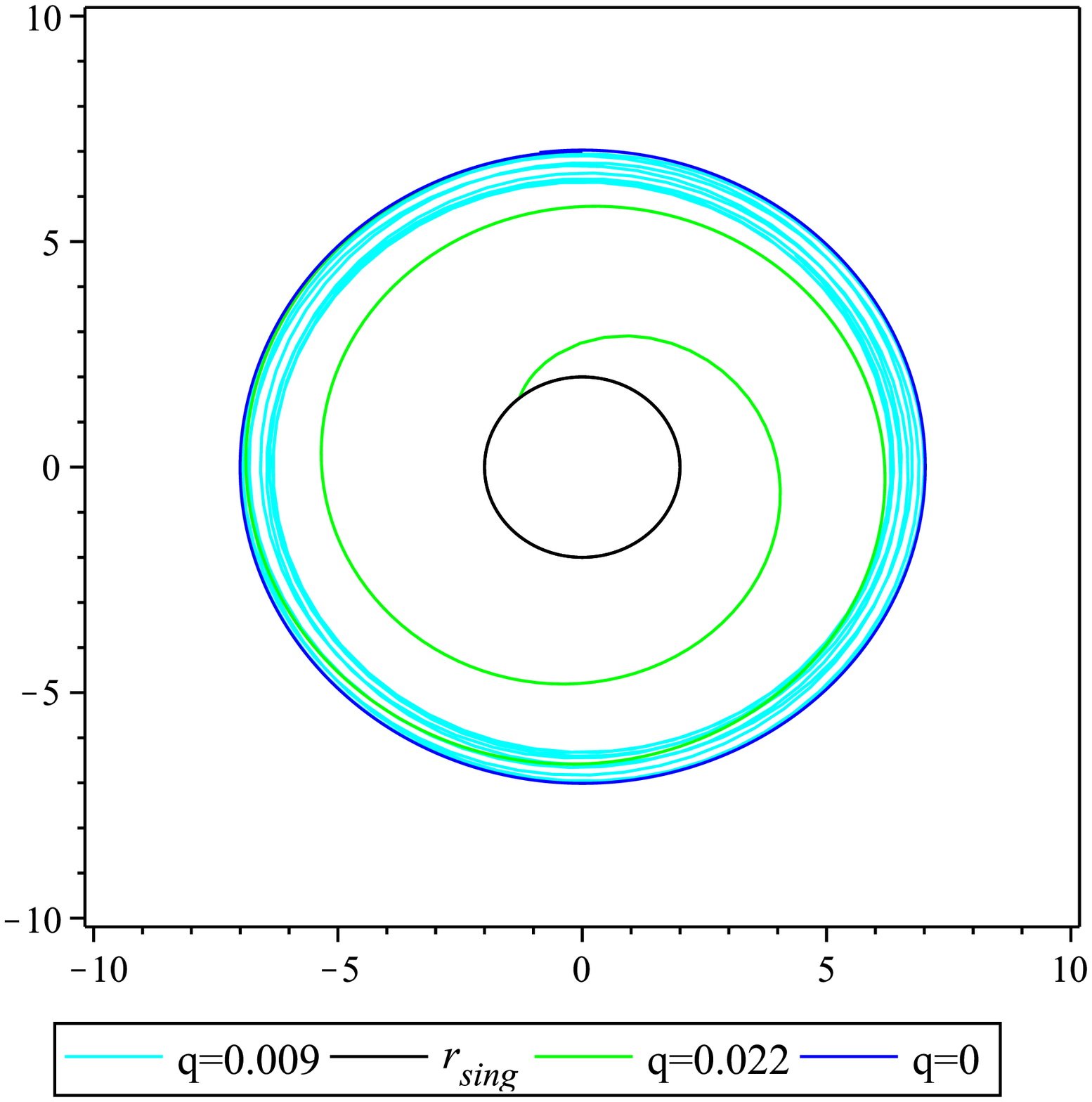}%
\quad
\includegraphics[scale=0.25]{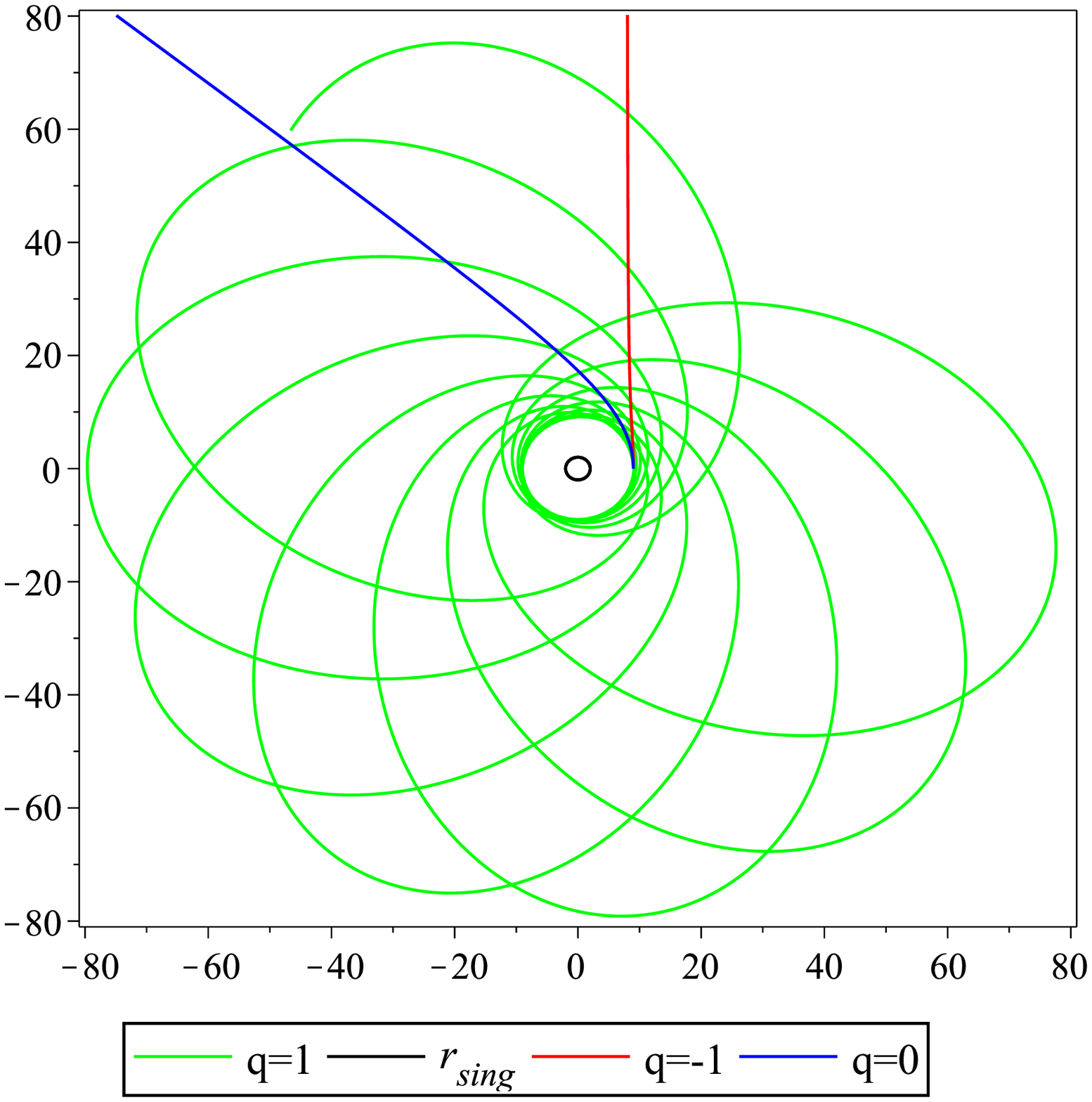}%
\quad
\includegraphics[scale=0.25]{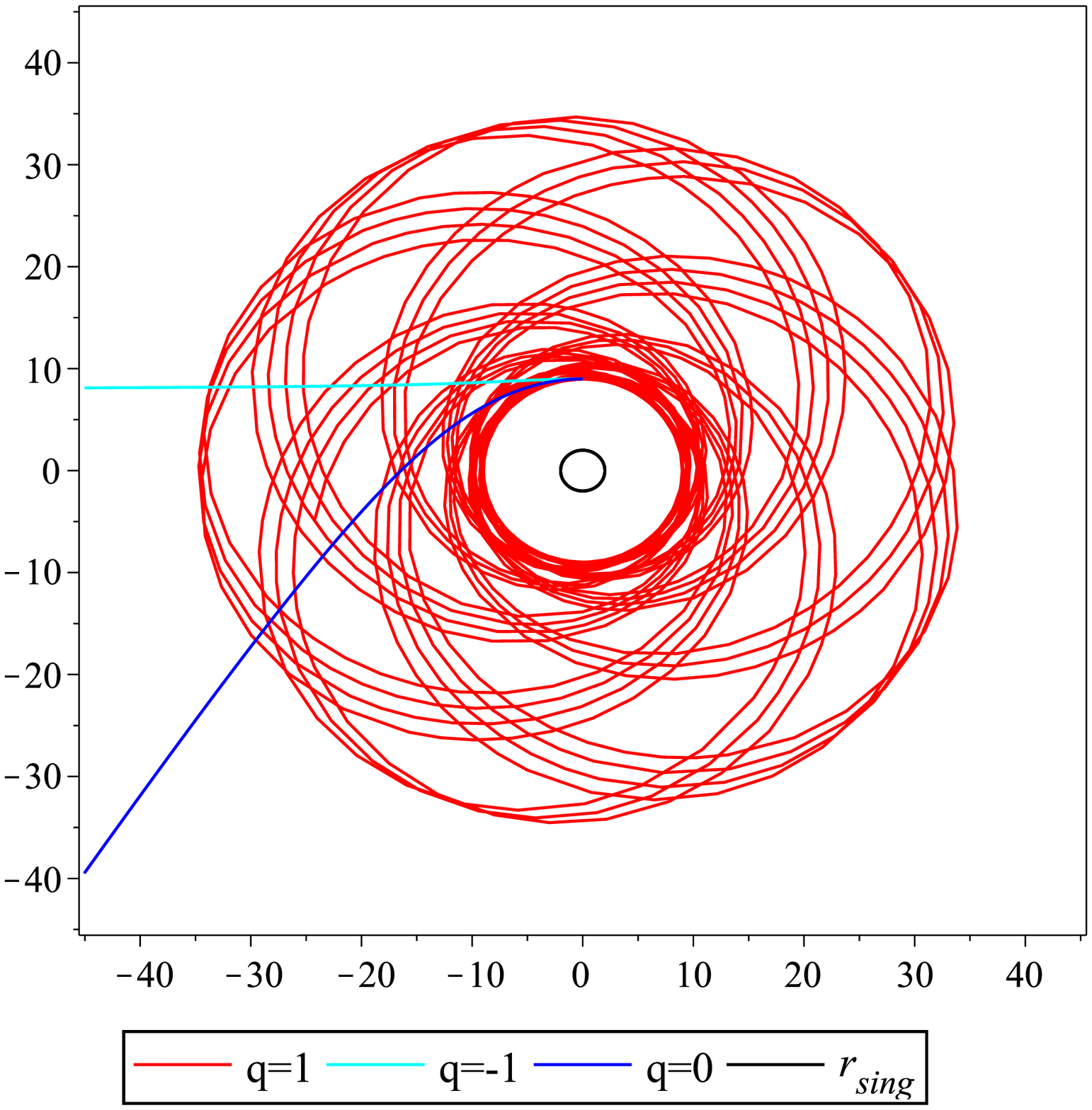}%
\caption{Influence of the quadrupole on unbounded Schwarzschild orbits with vanishing initial radial velocity ($\dot r(0)=0)$.
Initial conditions: 
$ \varphi(0) = \pi/2,  \ r(0) =7, \ \dot{\varphi}(0)= 0.07145$ (left plot);
$ \varphi(0)=0 $, $r(0)=9$, $\dot{\varphi}(0)=0.075$ (center plot);
$\varphi(0)=\pi /2$, $r(0)=9$, $\dot{\varphi}(0)=0.07145$ (right plot).}
\label{fig4b}%
\end{figure}

\begin{figure}%
\includegraphics[scale=0.25]{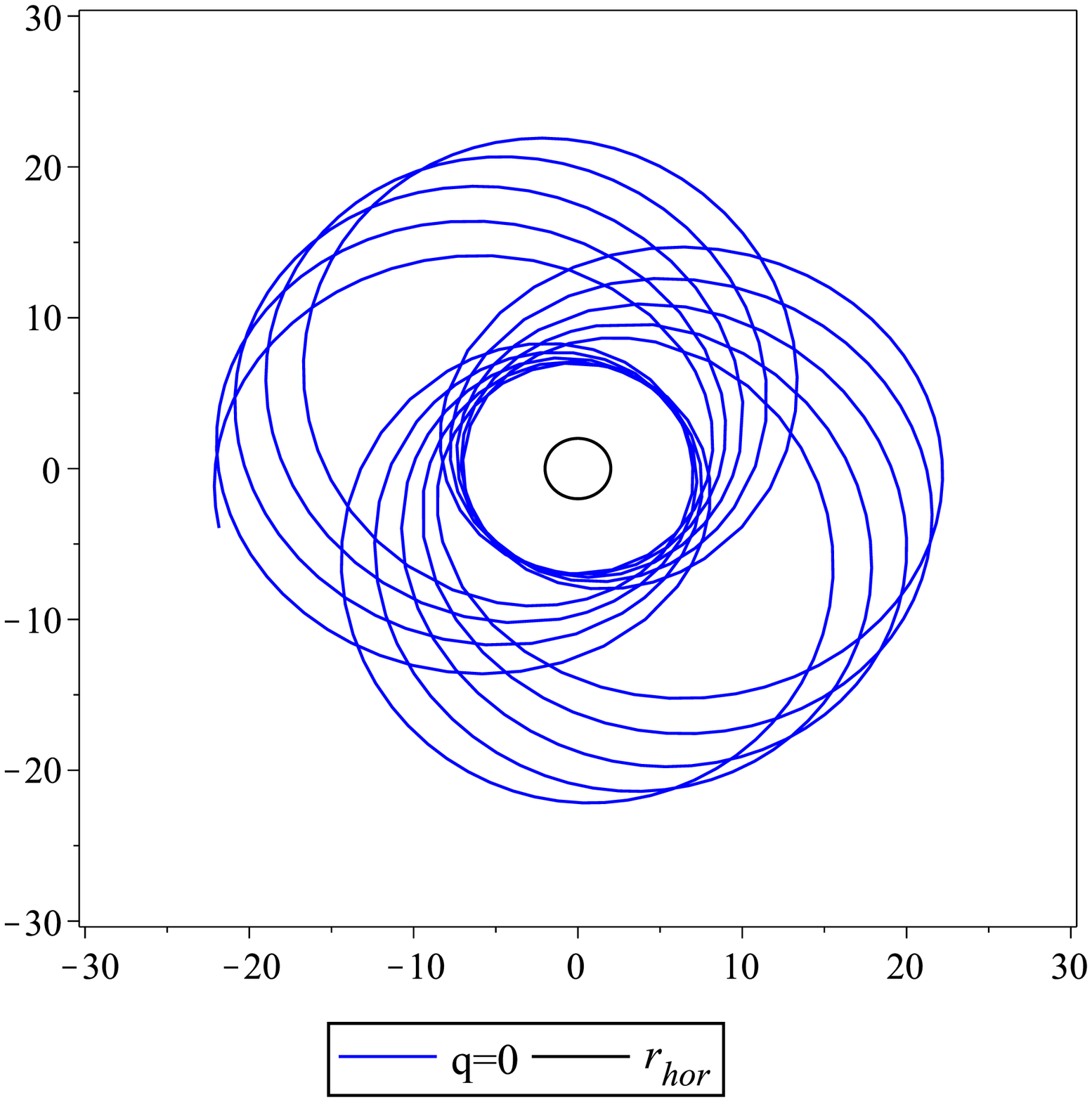}%
\quad
\includegraphics[scale=0.25]{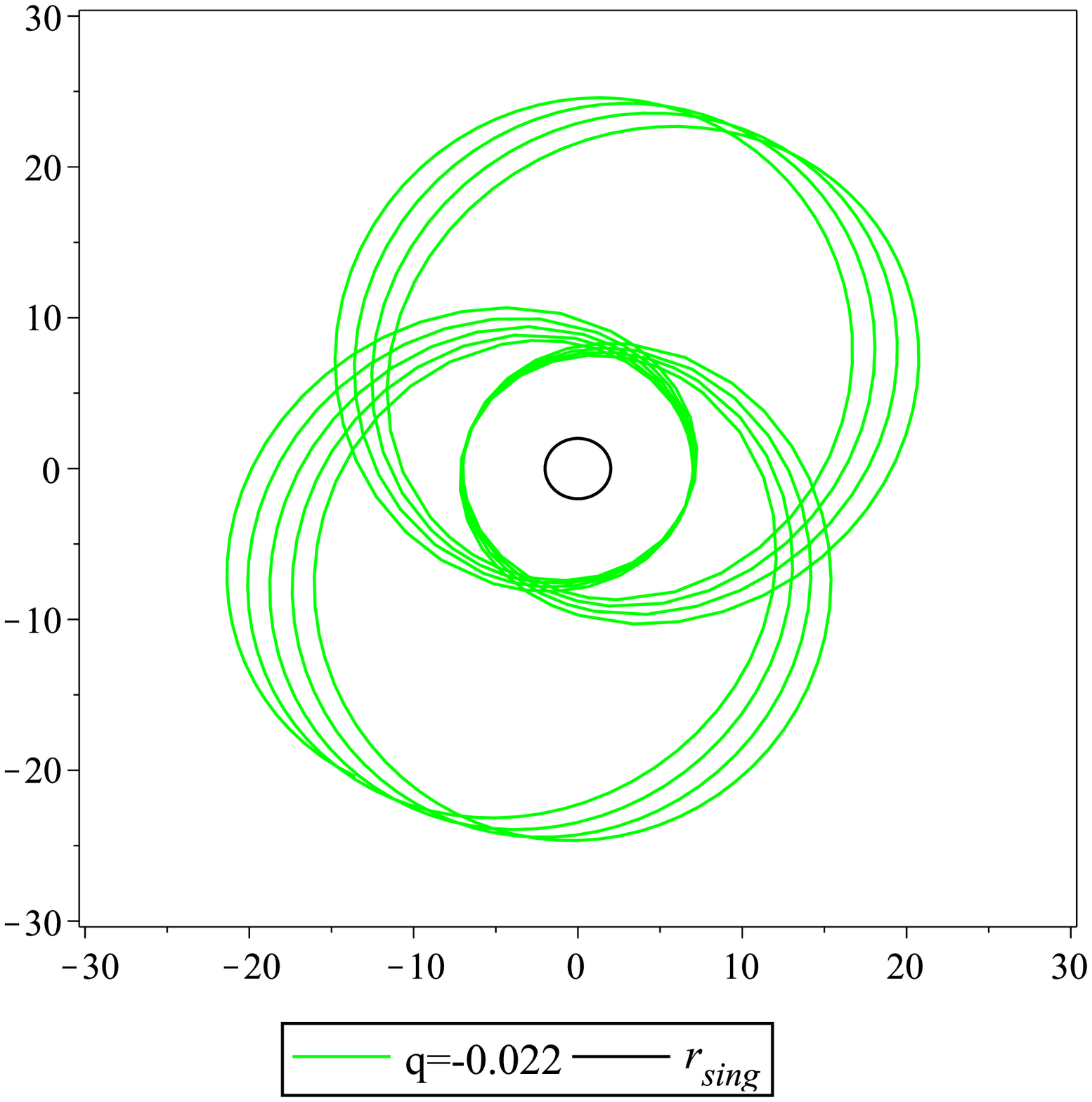}%
\quad
\includegraphics[scale=0.25]{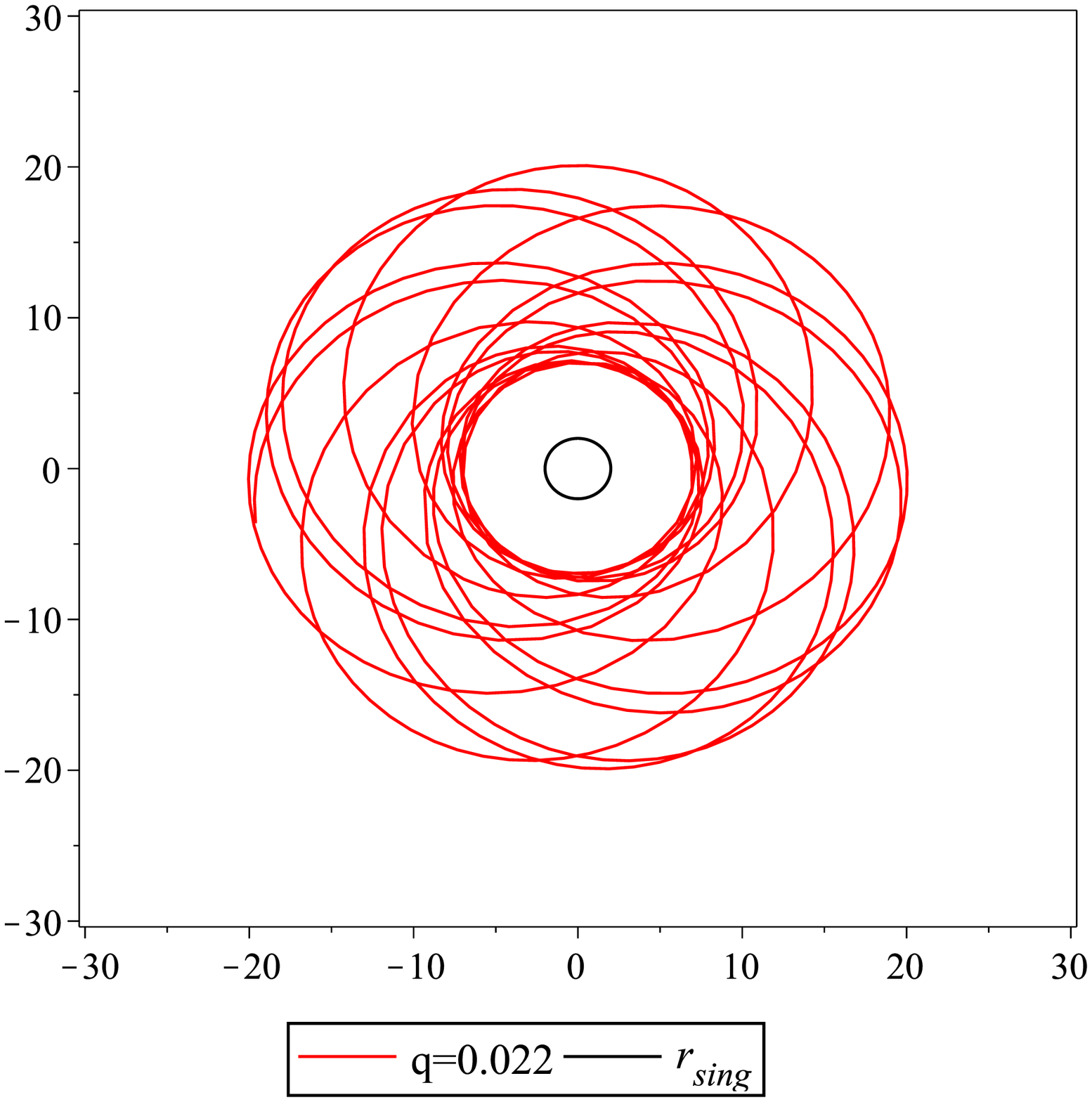}%
\caption{Influence of the quadrupole on Schwarzschild bounded orbits with vanishing initial radial velocity ($\dot r(0)=0)$.   
The initial conditions are  $\varphi(0)=0$, $r(0)=7$,     and $\dot{\varphi}(0)=0.08$ for all the trajectories. }
\label{fig4c}%
\end{figure}

As mentioned above the equatorial plane $\theta=\pi/2$ is a geodesic plane due to the symmetry of the gravitational source. In this case, the geodesic equations reduce to 
\label{t-eq}
\be
\dot{t} =   {E}\left(1-\frac{2m}{r}\right)^{-(1+q)},
\ee
\label{phi-eq}
\be
\dot{\varphi} = \frac{  {l}}{r^2}\left(1-\frac{2m}{r}\right)^{q} ,
\ee
\be
\left(1+\frac{m^2}{r^2-2mr}\right)^{-q(2+q)} \dot r ^2 =  {E}^2- \left(1-\frac{2m}{r}\right)^{q+1} \left[\frac{  l ^2}{r^2}\left(1-\frac{2m}{r}\right)^{q} +\epsilon \right] .
\label{eq-r}
\ee
To investigate the behavior of the geodesics it is convenient to introduce the new coordinate $u=1/r$ and use the azimuthal angle $\varphi$ instead of the affine parameter $\tau$ so that
\be
\frac{du}{d\tau} =  {  l }{u^2} \left(1-{2m}{u}\right) ^q \frac{du}{d\varphi} \ .
\ee
Then, the geodesic equation (\ref{eq-r})  can be expressed as
\be
\frac{ (1-2mu)^{q(4+q)} }{(1-mu)^{2q(2+q)}} \left(\frac{du}{d\varphi}\right)^2 = F(u,q)\ ,
\ee
where
\be
F(u,q) = \frac{1}{  l ^2} \left\{   E ^2 - (1-2mu)^{1+q} \left[   l ^2 u^2 (1-2mu)^q + \epsilon\right] \right\}\ .
\ee
The properties of the geodesic motion are then determined by the function $F(u,q)$ which must be positive for the velocity to be a real valued function. Furthermore, the zeros of $F(u,q)$ are the points where the radial velocity vanishes, i.e., the perihelion or aphelion distance in the case of bounded orbits. In turn, the behavior of $F(u,q)$ can be explored by comparison with the function $f(u)$ which is defined 
as 
\be
f(u)=F(u,q=0) = \frac{1}{  l ^2} \left[  E ^2 - (1-2mu) (  l ^2 u^2 + \epsilon)\right] \ .
\ee
This means that the function $f(u)$ determines the behavior of the geodesics in the case of the Schwarzschild spacetime and, consequently, a comparison between $F(u,q)$ and $f(u)$ will determine the main differences between geodesics of the $q-$metric and those of the Schwarzschild spacetime. It is easy to  show that 
\be
F(u,q) > f(u) \quad \hbox{for}\quad q>0\quad \hbox{and} \quad F(u,q) < f(u) \quad \hbox{for}\quad q<0
\ee
in the interval $u\in \left(0,\frac{1}{2m}\right)$ i.e., in the spatial region located between the outer singularity $(r=2m)$ and infinity. This behavior is illustrated in Fig. 
\ref{fig3}. 
This simple observation leads to an interesting physical result. Suppose that the zeros of $f(u)$ are $u_1$ and $u_2$ with $u_1<u_2$ so that the trajectory is bounded between 
$r_1=1/u_1$ and $r_2=1/u_2$; consequently, the perihelion distance is $r_2<r_1$. Suppose that the value of $q$ is so chosen that $F(u,q)$ 
has also two zeros at $u_1'<u_2'$. Then, from the behavior of $F(u,q)$ (cf. Fig. \ref{fig3}) we conclude that
\be
u_1 < u_1'\ , \quad \hbox{i. e.}\quad r_1> r_1'\quad \hbox{and} \quad u_2' < u_2 \ , \quad \hbox{i. e.}\quad r_2' > r_2\quad \hbox{for} \quad q>0 \ .
\ee
This means that the perihelion distance for positive $q$ is greater that the one for vanishing quadrupole. A similar analysis shows that the perihelion distance decreases for negative values of the quadrupole. We conclude that the presence of the quadrupole in the $q-$metric leads to a change of the perihelion distance, an effect that has been predicted also for other metrics with quadrupole moment \cite{qp89,quev90}. Some examples of geodesics for different values of the quadrupole parameter are given in Figs. \ref{fig4a}-\ref{fig4c}, where for simplicity the radial coordinate has been redefined as $r/m$ for all the trajectories.   

In Fig. \ref{fig4a}, we consider unbounded Schwarzschild orbits with non-vanishing initial radial velocities under the influence of the quadrupole.
We see that for the chosen initial radial and angular velocities all the test particles are able to escape from the gravitational field of 
the naked singularity.   
The value of the quadrupole determines only the direction along which the particle escapes towards infinity.  

In Fig. \ref{fig4b}, we investigate how the quadrupole acts on unbounded Schwarzschild orbits with zero initial radial velocity. 
First,  we study the change of a stable circular Schwarzschild geodesic under the influence of a quadrupole (left panel). For a small quadrupole 
($q=0.009$), the geodesic moves slowly towards the central singularity and finally reaches a stable circular orbit with a radius which is smaller than the initial Schwarzschild radius. In the central and right plots, we consider the same values for the quadrupole moment, but different initial angular velocities 
$\dot\varphi(0)$. In both cases, the test particles move along  bounded orbits with different perihelion shifts. From the behavior of these trajectories, we
conclude that the quadrupole is able to transform unbounded Schwarzschild trajectories into bounded trajectories.

In Fig. \ref{fig4c}, we consider a Schwarzschild bounded orbit under the influence of the quadrupole  with zero initial radial velocity and the same non-zero
value for the initial angular velocity. 
The left panel shows the Schwarzschild geodesic, whereas the central and right plots are for a particular negative and positive value of the quadrupole, respectively.
In this case, we see that the quadrupole does not affect the bounded character of the geodesic. However, it can drastically modify the geometric structure 
of the trajectory.  

In general, we see that the quadrupole of the naked singularity always affects the motion of test particles. The specific change can manifest itself 
in different ways, depending on the explicit value of the quadrupole.

\subsection{Circular orbits}
\label{sec:cio}

\begin{figure}%
\includegraphics[scale=0.4]{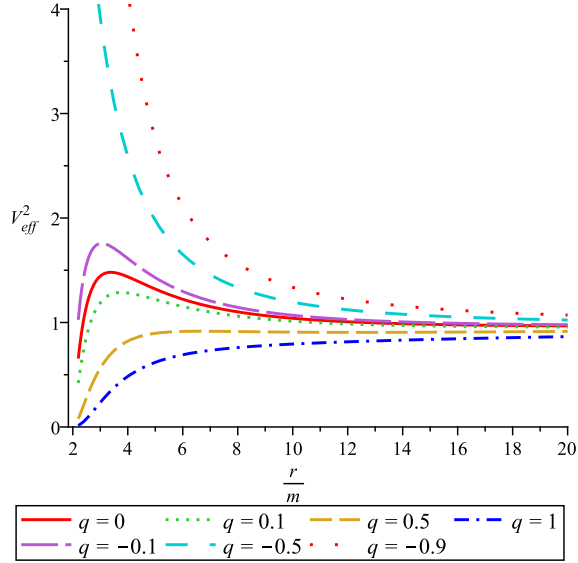}%
\caption{The effective potential for timelike circular geodesics on the equatorial plane as a function of the radius for different values of the quadrupole parameter. Here we set $l^2=30$ for concreteness. }%
\label{fig5}%
\end{figure}

Circular orbits on the equatorial plane can be investigated analytically. In fact, in this case the motion under the condition $\dot r=0$ is equivalent to the motion of a test particle in the effective potential [cf. Eq.(\ref{eq-r})]
\be
\label{veff}
V_{eff}^2(r,q)=  \left(1-\frac{2m}{r}\right)^{q+1} \left[\frac{  l ^2}{r^2}\left(1-\frac{2m}{r}\right)^{q} +\epsilon \right].
\ee
The presence of the quadrupole parameter influences the behavior of the effective potential, as can be seen in Fig. \ref{fig5}.
The effective potential of the Schwarzschild is also shown for comparison. For positive values of $q$ the value of the effective potential at a given point outside the outer singularity is always less than the Schwarzschild value. For negative values of the quadrupole the opposite is true. This indicates that the distribution of circular orbits on the equator of the $q-$metric can be modified drastically by the quadrupole. We will now explore this point in detail. 

The explicit value of the angular momentum of a test particle along a circular orbit can be derived from the condition $\partial V_{eff}^2 /\partial r =0$. A straightforward computation yields that this condition is satisfied if the angular momentum is given by
\label{ang-moment}
\be
\label{l2co}
 { l} ^2=\frac{m(q+1)\left(1-\frac{2m}{r}\right)^{-q} r^2}{r-(3+2q)m} ,
\ee
an expression from which it follows immediately that the radius of any circular orbit must satisfy the condition 
\be
r_c> m(3+2q) \ .
\label{acr}
\ee
This radius is greater than the Schwarzschild radius for a positive quadrupole parameter. However, for negative values of $q$ the radius is smaller than the Schwarzschild value and, in principle, can be made as close as possible to the outer singularity located at $r=2m$. Below we will show that in the limiting case 
$r_c=m(3+2q)$, the orbit becomes lightlike.

Furthermore, the energy of a test particle in circular motion can be expressed as
\be
 {E} ^2= \left(1-\frac{2m}{r}\right)^{q+1}\left[ \frac{m(q+1)}{r-m(3+2q)}+1\right] \ ,
\ee
which is positive only within the range of allowed radii (\ref{acr}). To completely characterize the parameters of circular orbits, we also calculate their angular velocity $\Omega(r)$ and period $T(r)$ and obtain
\be
\Omega(r)= \dot\varphi = \frac{1}{r} \left[\frac{m(1+q) \left(1-\frac{2m}{r}\right)^q  }{ r- m(3+2q) }\right]^{1/2}\ ,
\ee
\be
T(r) = \int \frac{\dot t}{\dot\varphi} d\varphi = 2\pi \frac{dt}{d\varphi} = 2\pi r^{3/2} \left[ \frac{r-m(2+q)}{m(1+q)(r-2m) }\right]^{1/2} 
\left(1-\frac{2m}{r}\right) ^{-q}\ ,
\ee
respectively.


\subsection{Stability analysis}
\label{sec:sta}

\begin{figure}%
\includegraphics[scale=0.3]{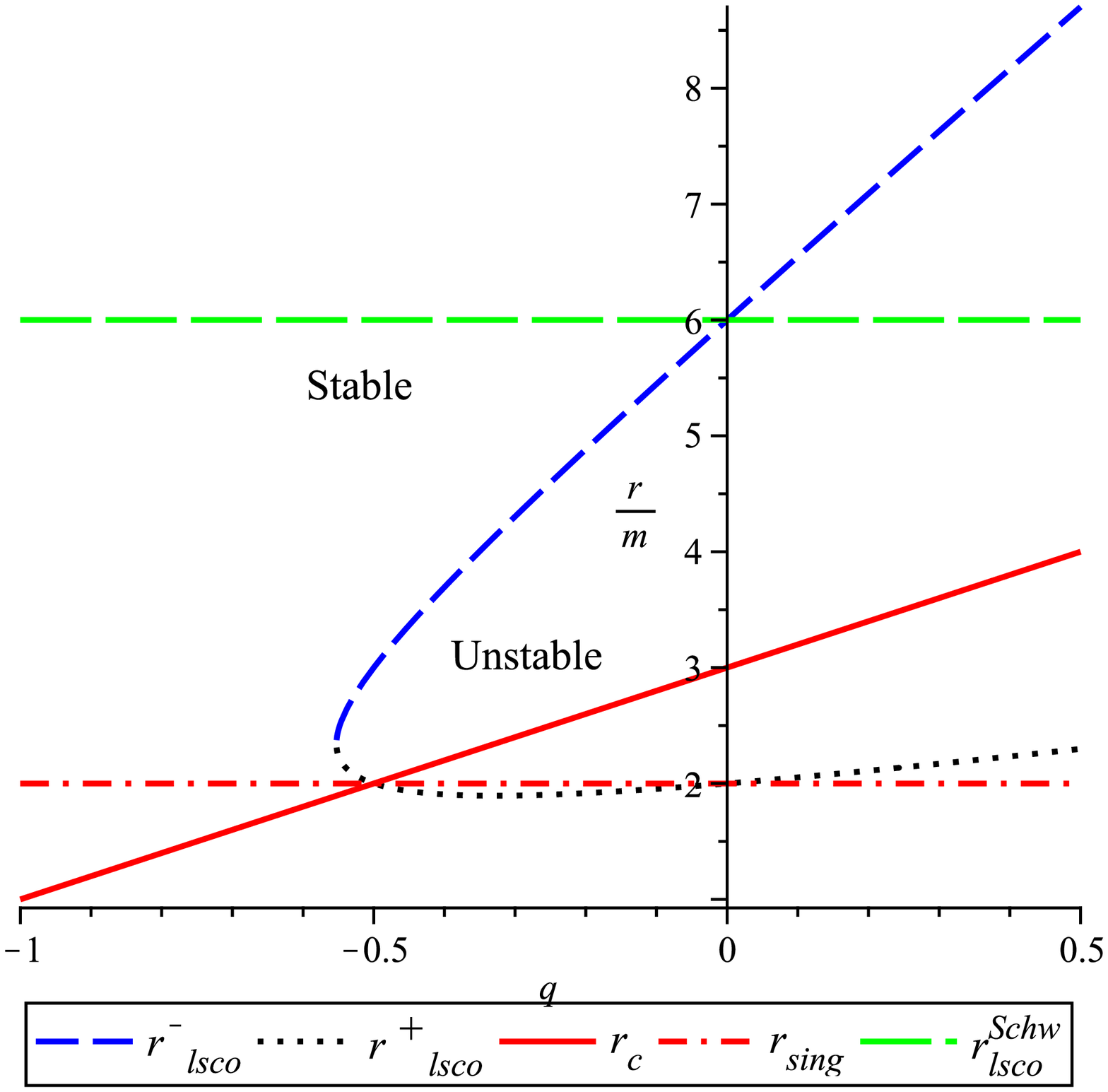}%
\qquad
\includegraphics[scale=0.3]{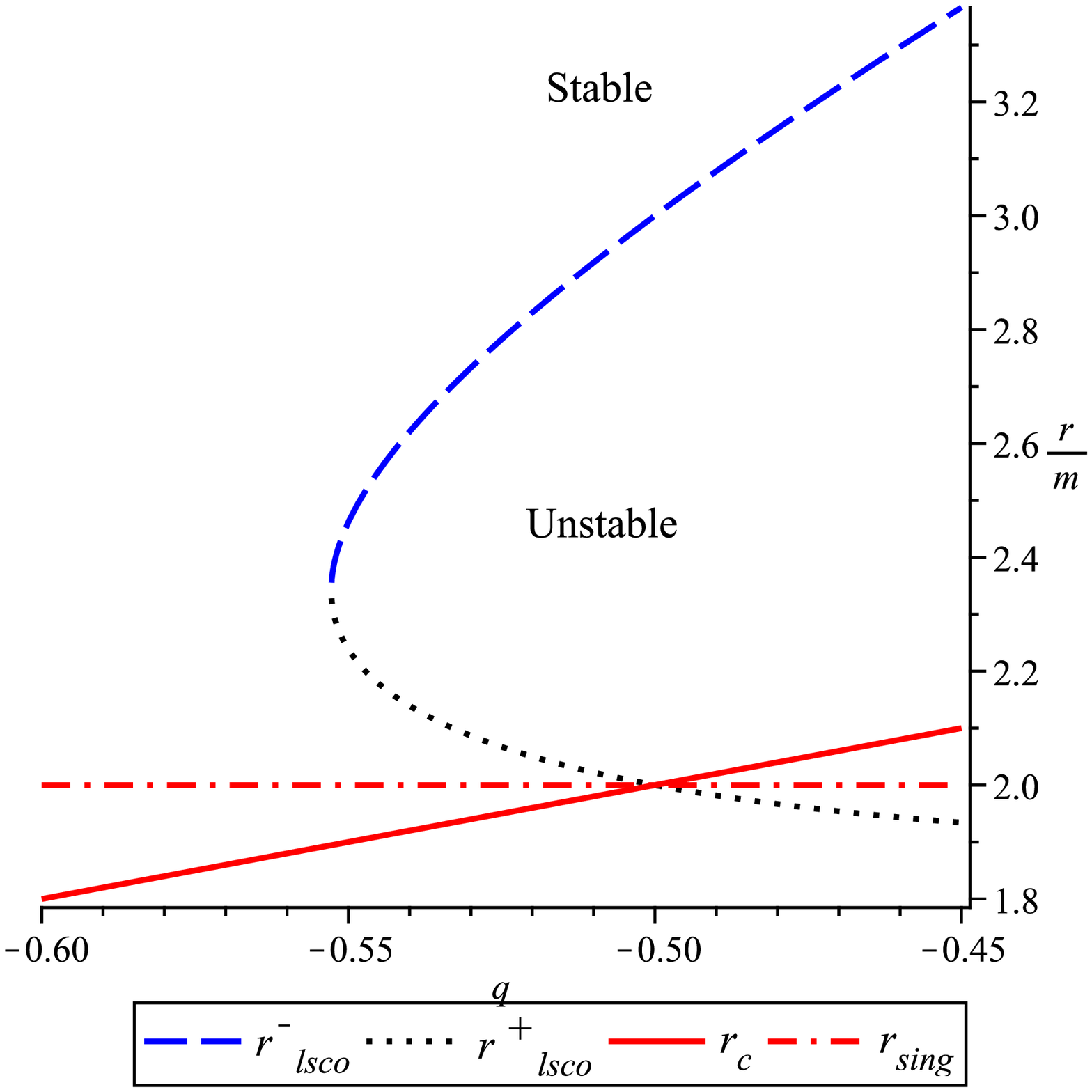}
\caption{The last stable circular radius for a timelike particle as a function of the quadrupole parameter $q$.  
The critical radius $r_c=m(3+2q)$ and the outer singularity $r_{sing}=2m$ are also plotted. The left plot is a zoom of the region of intersection.}
\label{fig6}%
\end{figure}

We now analyze the stability properties of the circular motion. In particular the last stable circular orbit is determined by the inflection points of the effective potential (\ref{veff}), i.e., by the zeros of the equation
\be
\frac{\partial^2 V_{eff}^2}{\partial r^2} = 
2\, \left( 1-\frac{2m}{r}  \right) ^{q-1} \frac{ m \left( 1+q \right)  
\left( {r}^{2}-8\,m r -6\,mqr +  4\,{m}^{2}{q}^{2}+14\,{m}^{2}q+ 12\,{m}^{2} \right) } 
{{r}^{4} \left[ r- m(3+2q) \right] } 
\ee
where we have replaced the value for the angular momentum of the circular orbit (\ref{l2co}). Then, we obtain the radii
\be
r_{lsco}^\pm = m(4 + 3 q \pm \sqrt{5q^2+10q + 4})
\ee
which are positive and lie outside the outer singularity for all allowed values of $q$. The behavior of $r_{lsco}^\pm$ is plotted in Fig. \ref{fig6} where we also plot the radius $r_c$ that denotes the minimum value at which the angular momentum $l^2$ is positive. 
It follows from the graph that 
it is necessary to consider three different intervals: I for $q\in (\infty, -0.5]$, II for $q \in (-0.5,-1+1/\sqrt{5}]$ and III for $q \in (-1+1/\sqrt{5},-1)$. The particular values of $q$ used to determine the different intervals follow from the conditions $r_-=r_c$ ($q=-0.5)$ and $r_+=r_-$ ($q=-1+1/\sqrt{5} \approx -0.5527)$.

In the interval I, the only valid radius is $r_{lsco}^+$ which is greater (smaller)  than the Schwarzschild radius for positive (negative) values of the quadrupole 
parameter $q$. Notice that on the boundary of this interval $(q=-0.5)$ the zero at $r_{lsco}^-$ cannot be considered because it is located on the outer singularity.
In this interval we have that ${\partial^2 V_{eff}^2}/{\partial r^2}>0$ so that all the circular orbits are stable. 
 If we imagine a hypothetical accretion disk made of test particles around the central source so that the inner radius of the disk coincides with 
$r_{lsco}^+$, the role of the quadrupole in the interval I consists in changing the inner radius. From an observational point of view, this implies that by measuring the inner radius of the disk, one can determine the value of the quadrupole. The smallest disk inner radius corresponds to the value $r_{lsco}^+(q=-0.5)=3m$ whereas the outer radius can in principle be extended to infinity. 

\begin{figure}%
\includegraphics[scale=0.3]{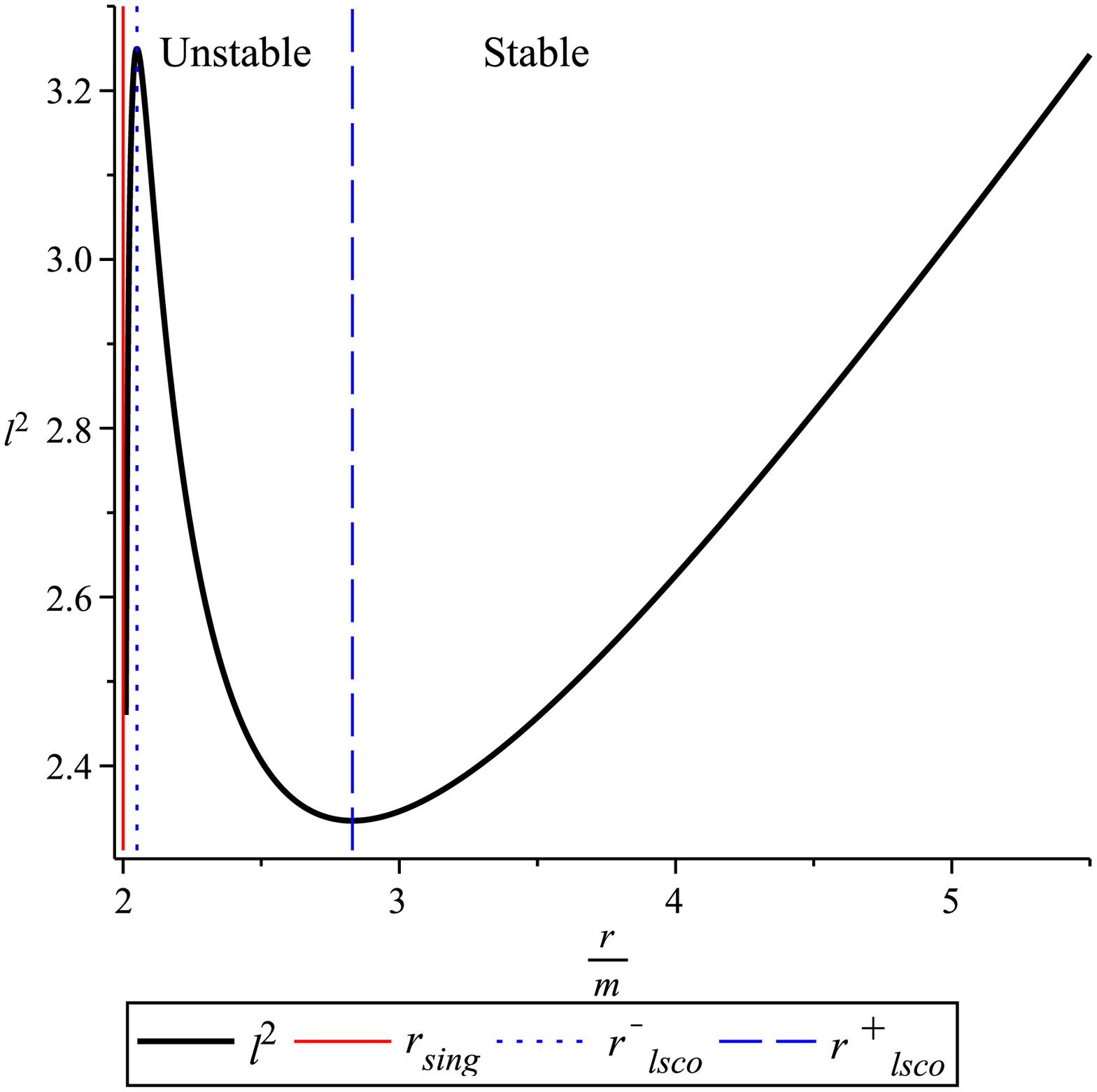}%
\qquad
\includegraphics[scale=0.3]{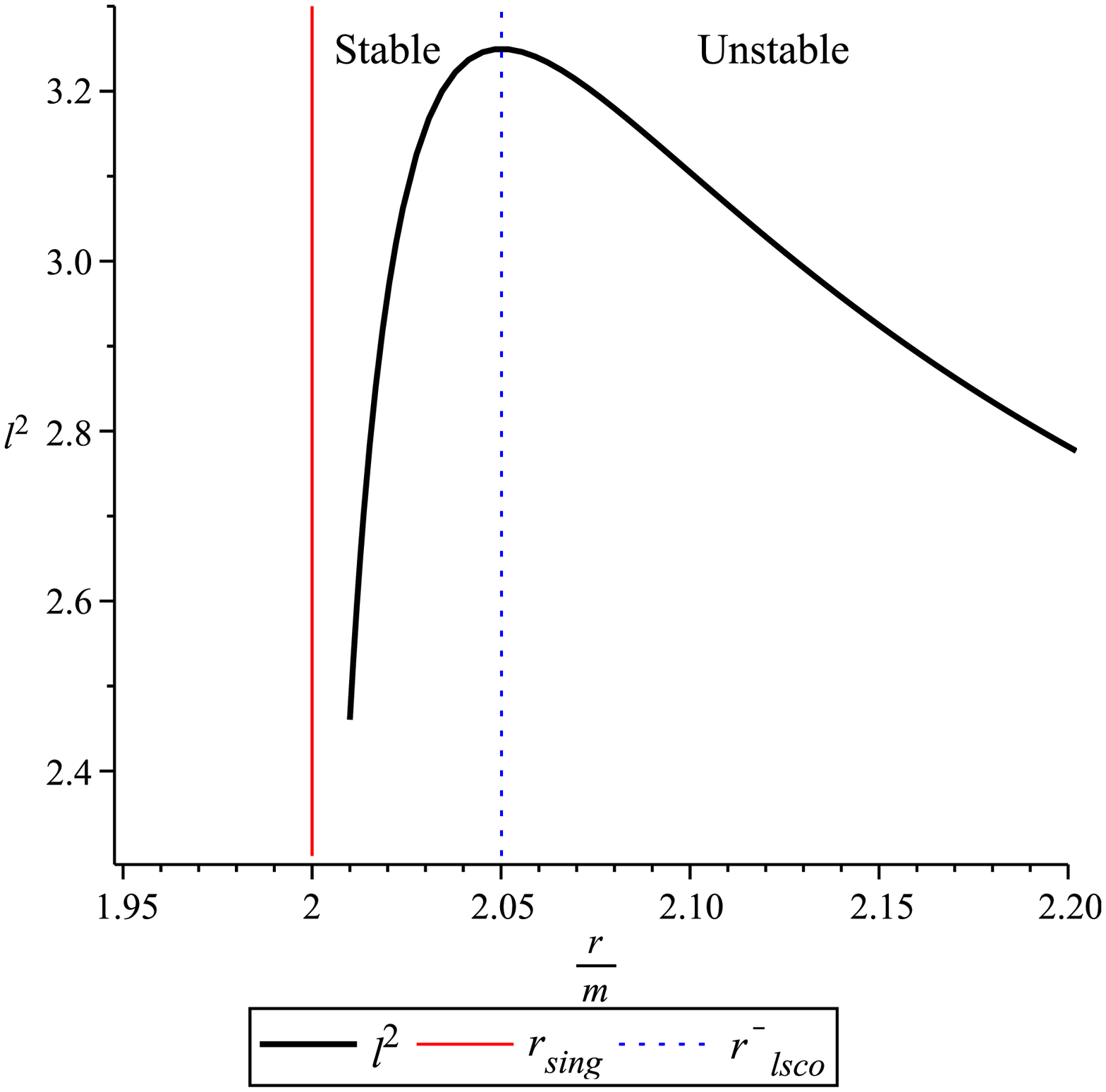}%
\caption{The angular momentum of test particles around a naked singularity with $q=-0.52$. The left plot is a zoom of the region where the internal stable ring is located.}%
\label{fig7}%
\end{figure}

In the interval II with $q\in (-0.5,-1+1/\sqrt{5})$, the second derivative of $V^2_{eff}$ is negative within the two zeros located at $r_{lsco}^+$ and $r_{lsco}^-$. This means that circular orbits are unstable within this interval. The radial extension $L$ of the unstable region is 
\be
L= r_{lsco}^+-r_{lsco}^- = 2m \sqrt{5q^2+10q + 4} = 
\left\{ \begin{array}{ll} m & \hbox{for } q = - 0.5 \\
                          0 & \hbox{for }  q = -1 + 1/\sqrt{5} 
\end{array} \right.
\ee
The interesting fact in this case is that there exists an additional stable region located between the outer singularity and $r_{lsco}^-$ (cf. Fig. \ref{fig6}) 
which is separated by the unstable region ($r_{lsco}^-,r_{lsco}^+$) from the exterior stable region with $r\geq r_{lsco}^+$. 
An accretion disk around such an object would consist of an external disk which can extend from $r_{lsco}^+$ to infinity and an internal ring located inside the 
region $(2m, m+3m/\sqrt{5}\approx 2.34m)$, where the outer boundary corresponds to the point where 
$r_{lsco}^+=r_{lsco}^-$. Since the internal stable ring is located so close to the outer singularity one could imagine that the angular momentum of the test particles should be very high in order to maintain the orbit. A detailed analysis, however, shows that this is not true. In Fig. \ref{fig7}, we plot the value of the angular momentum as a function of the radial distance for a particular quadrupole parameter which 
allows the existence of an internal ring. 
We can see that the maximum value of the angular momentum is reached on the external boundary of the ring and then it decreases as the singularity is approached. The explicit value of the angular momentum is not high and in fact it is comparable with the values for test particles located inside the external stable disk at distances of several times $m$. The peculiarity of the angular momentum is that its value decreases  as the singularity is approached, an effect that contradicts the physical
expectations for an attractive gravitational field.
 The only possible explanation for this unusual behavior is that there is an additional force that compensates the gravitational attraction. Similar effects has been found in the gravitational field of 
other naked singularities and have been interpreted as a manifestation of repulsive gravity \cite{pqr11a,pqr11b,pqr13b,pq15}. An alternative approach in which repulsive gravity is defined in terms of curvature invariants leads to similar results \cite{lq12,lq14}. 

In the interval III, with $q \in (-1+1/\sqrt{5},-1)$, there are no last stable orbits. All the circular trajectories are allowed, in principle, starting 
at the outer singularity $r=2m$. An analysis of the angular momentum shows that it diminishes as the naked singularity is approached. Again, the simplest
explanation is to assume the presence of repulsive gravity. An accretion disk around this type of naked singularities would have a continuous structure which extends from
the singularity to infinity.

\subsection{Circular null geodesics}
\label{sec:cng}

In the case of null geodesics ($\epsilon=0)$, the effective potential (\ref{veff}) reduces to 
\be
V_{eff}^2 = \frac{l^2}{r^2} \left(1-\frac{2m}{r}\right)^{1+2q}
\ee
and its behavior for a given $l^2$ is similar to that shown in Fig. \ref{fig5}. The important quantities for the analysis of circular motion of photons are
\be
\frac{\partial V_{eff}^2 }{\partial r} = - \frac{2l^2}{r^4} \left(1-\frac{2m}{r}\right)^{2q}[r-m(3+2q)]\ ,
\ee
and
\be
\frac{\partial^2  V_{eff}^2 }{\partial r^2} = \frac{2l^2}{r^6}\left(1-\frac{2m}{r}\right)^{1+2q} [3r^2-12qmr-18mr+28qm^2+8q^2m^2+24m^2]\ .
\ee
The first derivative vanishes for $r=r_\gamma=m(3+2q)$, 
an expression which reduces to the standard Schwarzschild value for $q=0$, as expected. We see that the circular 
orbit located at $r_\gamma=m(3+2q)$ corresponds to the trajectory of a photon.  This is also the limiting radius for timelike circular geodesics, as explained above.

By inserting this radius value into the second derivative of the effective potential, we
obtain an expression which is negative for all values of $q$. This implies that the circular orbit with radius 
$r_\gamma=m(3+2q)$ is unstable, independently of the value of the angular momentum. 
In contrast to the Schwarzschild spacetime, where there is only one circular orbit with $r_\gamma=3m$, in the case of the $q-$metric the radius $r_\gamma=m(3+2q)$ can take any value within the interval $(2m,\infty)$, depending on the value of the quadrupole parameter. Nevertheless, for fixed values of $m$ and $q$ only one circular orbit is allowed.


\section{Radial geodesics}
\label{sec:rad}

\begin{figure}%
\includegraphics[scale=0.7]{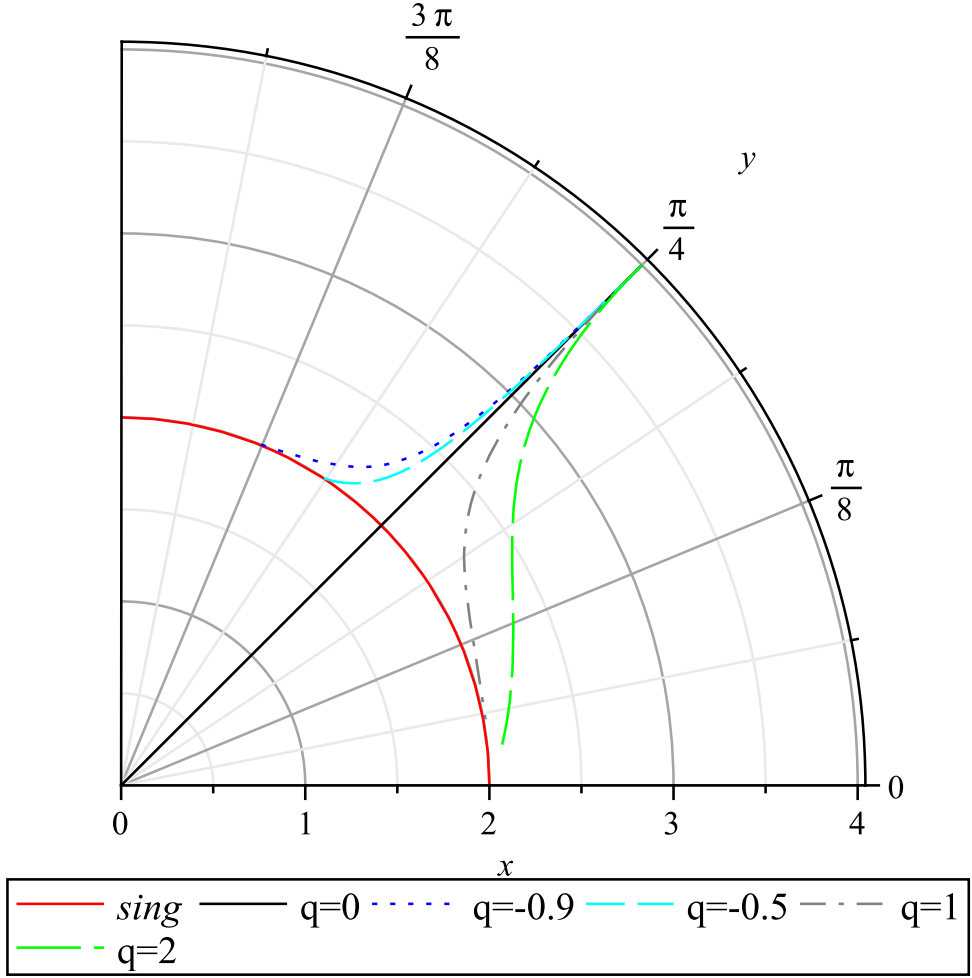}%
\caption{Free fall of test particles with initial point located at $r=4\ (m=1)$ and $\theta=\frac{\pi}{4}$. The singularity at $r=2m$ and the radial geodesic for the 
Schwarzschild spacetime are also shown for comparison. }%
\label{fig8}%
\end{figure}

In the Schwarzschild spacetime, radial geodesics representing the free fall of test particles are straight lines that connect the initial point in spacetime with the singularity located at the origin of coordinates. Therefore, it is useful to study radial geodesics in the spacetime described by the $q-$metric in order to evaluate 
the influence of the quadrupole. We will consider the free fall of test particles with vanishing angular momentum ($l=0$). Then, the motion along the radial coordinate is governed by the equation
\be
\dot r = -\left( 1 + \frac{m^2 \sin^2{\theta}}{r^2 - 2 m r}  \right)^{\frac{q}{2}(2 + q)}\sqrt{E^2-\Phi_r^2} \ ,
\label{eqrr}
\ee
where the minus sign indicates that the particle falls inward and
\be
\Phi^2_r =  \left( 1 -\frac{2m}{r} \right)^{1 + q} \left[   r^2 \left( 1 -\frac{2m}{r} \right)^{- q} \left( 1 + \frac{m^2 \sin^2\theta}{r^2 - 2m r}\right)^{-q(2     +   q)} {\dot\theta}^2 +  \epsilon \right] \ .
\label{eqthetar}
\ee
In addition, we must take into account that an arbitrary polar plane ($\theta=$const) is not necessarily a geodesic plane. This means that the radial geodesic is determined by Eq.(\ref{eqrr}) together with the equation (\ref{ddtheta}) for the evolution along the angle $\theta$. We illustrate the result of the integration of these two differential equations in Fig. \ref{fig8}.
The effect of the quadrupole parameter $q$ becomes plausible from the behavior of the geodesics. For sources with negative quadrupole parameter (prolate sources), the geodesics deviate from the initial angle $\theta=\frac{\pi}{4}$ towards the axis of symmetry. In the case of positive quadrupole $q$ (oblate body), the deviation is towards the equatorial plane of source. This result is in accordance with our physical expectations. Indeed, an oblate source is larger on the equatorial plane and so one would expect that the mass density is larger on the equator, generating a larger gravitational interaction. In the case of prolate bodies, the same intuitive reasoning is valid for the symmetry axis. This result reinforces our interpretation of the parameter $q$ as determining the quadrupole moment of the gravitational source. 

In the case of radial geodesics, it is interesting to compute the coordinate time, $t$, and the proper time, $\tau$, that passes along the trajectory of 
a test mass between two radii $r_i$ and $r_f$. For simplicity, we consider the case of radial geodesics on the equatorial plane where the geodesic equations
reduce to 
\be
\dot r ^2 = \left(1 + \frac{m^2}{r^2-2m}\right)^{q(2+q)} \left[ E^2 -\left(1-\frac{2m}{r}\right)^{1+q}\right]  \ , 
\quad \dot t = E \left(1-\frac{2m}{r}\right)^{-1-q} \ .
 \ee
It is then easy to obtain the relationships
\label{del-Tau}
\be
\Delta \tau = \int^{r_f} _{r_i}  \left[ {E}^2-\left(1-\frac{2m}{r}\right)^{q+1}\right]^{-1/2}\left( 1+\frac{m^2}{r^2-2mr}\right)^{-\frac{1}{2}q(2+q)} dr \ ,
\ee
and 
\be
\Delta t = \int^{r_f} _{r_i}  {E}\left( 1-\frac{2m}{r}\right)^{-(1+q)} \left[ {E}^2-\left(1-\frac{2m}{r}\right)^{q+1}\right]^{-1/2}\left( 1+\frac{m^2}{r^2-2mr}\right)^{-\frac{1}{2}q(2+q)} dr\ ,
\label{del-t}
\ee
for the proper time and coordinate time, respectively. Let us consider the case of a particle located initially at $r_i\rightarrow\infty$ so that $E^2 =1$. If the final 
state is at the singularity $r_f=2m$, we obtain
\be
\tau_{2m} =  \int_{\infty}^{2m} \left(\frac{2m}{r}\right)^{-\frac{1}{2}(1+q)} \left(1+\frac{m^2}{r^2-2mr} \right)^{-\frac{q}{2}(2+q)} dr \ .
\ee
It is not possible to find an analytical expression for this integral. Nevertheless, one can evaluate it either numerically or by means of a Taylor expansion around
the value $q=0$ and $r=2m$. As a result we obtain that the proper time is finite for all the allowed values of $q$. The same result is obtained for the Schwarzschild spacetime $(q=0)$. This results indicates that the quadrupole does not affect the finiteness of the proper time that is necessary to reach the hypersurface $r=2m$.

In a similar manner, for the coordinate time we obtain
\be
t_{2m} = E \int_\infty^{2m} \left[ E^2 - \left( 1-\frac{2m}{r}\right) ^{1+q}\right]^{-1/2} \left( 1-\frac{2m}{r}\right)^{-1+\frac{q^2}{2}} 
\left(1-\frac{2m}{r} + \frac{m^2}{r^2}\right)^{-\frac{q}{2}(2+q)} \ .
\ee 
Again, a Taylor expansion of this quantity reveals that it is finite for $q<0$ and infinite for $q\geq 0$. This means that a positive quadrupole does not change 
the main property of the coordinate time of the Schwarzschild spacetime. However, in the case of negative quadrupole the coordinate time completely changes so that an observer at infinity would observe how the test particle reaches the outer singularity located at $r=2m$.

Let us now consider the radial motion of photons. Suppose that a photon is emitted at a radius $r_{em}$ with frequency $\nu_{em}$ and received at a radius $r_{rec}$ with frequency $\nu_{rec}$. Then, the redshit $z$ in a static spacetime is determined by the relationship \cite{wald}
\be
z+1 = \frac{\nu_{em}}{\nu_{rec}} = \left(\frac{g_{tt}|_{rec}}{g_{tt}|_{em}}\right)^{1/2} = \left(\frac{1-2m/r_{rec}}{1-2m/r_{em}}\right)^{\frac{1}{2}(1+q)} \ .
\ee
If the emission occurs near the outer singularity ($r_{em} \sim 2m)$, the redshift diverges, independently of the value of $q$. This means that for an observer located outside the radius $r=2m$, it is not possible to receive information from the singularity. In this sense, the singularity remains invisible for external observers.


\section{Conclusions}
\label{sec:con}

In this work, we investigated the motion of test particles in the gravitational field described by  the $q-$metric, the simplest generalization of the 
Schwarzschild metric which contains a quadrupole parameter.   It was shown that the $q-$spacetime possesses two singularities at $r=0$ and $r=2m$ which are not
covered by a horizon, i.e., they are not isolated from the exterior spacetime. 
For certain values of the parameter $q$, a third $\theta-$dependent singularity appears inside the spatial region contained within the radial interval $(0,2m)$. 
If we limit ourselves to the region with $r>2m$, the geodesic equations determine the motion of test particles around the naked singularity located at 
$r=2m$. First, we studied the bounded motion on an arbitrary polar plane $\theta\neq 0,\pi/2$ and showed that in general the plane of the orbit moves towards the
equator or to the axis of symmetry, depending on the value of the quadrupole. We performed a brief analysis of the conditions under which circular orbits are allowed for arbitrary values of the polar angle $\theta$, the angular momentum of the particle and the quadrupole parameter. 

We studied in detail the motion on the equatorial plane $\theta=\pi/2$. We integrated numerically the geodesics equations and found bounded and unbounded orbits for different values of the quadrupole. It was also shown explicitly that the presence of the quadrupole leads to a change of the perihelion distance of a bounded orbit.
This effect could in principle be used to determine the quadrupole moment of the gravitational source.

The analysis of circular orbits and their stability leads to  a classification of naked singularities according to which in certain cases 
it is possible to determine the quadrupole parameter by measuring the inner radius of the accretion disks made of test particles only.
 Suppose, for instance, that an accretion disk is detected with inner radius $r_{inner}$ within the interval ($3m,\infty$). Then, the naked singularity is of type III with a quadrupole parameter that can be determined numerically by using the formula 
$r_{inner}= m(4+3q + \sqrt{5q^2+10q+4})$ and the result must be a value contained within the interval $q\in (-0.5,\infty)$. 
If the value of the accretion disk inner radius is inside the interval $(2.34m,3m)$, the naked singularity belongs to the class II, 
and the quadrupole parameter can be
determined again from the formula $r_{inner}= m(4+3q + \sqrt{5q^2+10q+4})$. In this case, the quadrupole parameter 
must be contained within the interval $(-0.5527, -0.5)$, and there must exist an additional inner ring located inside the spatial region $(2m, 2.34m)$, whose exterior radius can be determined by the formula 
$r_{ext}= m(4+3q -\sqrt{5q^2+10q+4})$. Finally, if the inner disk radius is inside the interval ($2m,2.34m)$, the naked singularity is of class I with 
$q\in (-1,-0.5527)$. In the case of class III and II naked singularities, it is possible to find  the exact value of the quadrupole parameter by measuring the
inner radius of the accretion disk. This is not true in  the case of class I singularities, because all the circular orbits are stable in this region. However, 
if we could measure the radius and the angular momentum of a particular circular orbit located inside the region  ($2m,2.34m)$,  the quadrupole parameter 
could be determined by using the expression (\ref{l2co}) for the angular momentum. In this manner, we see that it is always possible to determine the quadrupole 
parameter from the physical properties of the accretion disk. The fact that in the gravitational field of naked singularities there exist circular orbits near the outer singularity is explained by assuming the presence of repulsive gravity. 

The study of radial geodesics starting at an arbitrary polar angle $\theta$ shows that the quadrupole induces a deviation from the original radial direction  
in such a way that the test particles tends either towards the equator or to the axis of symmetry, depending on the value of the quadrupole. We also calculated the
proper time and the coordinate time along radial geodesics on the equatorial plane and established the analogies and differences with respect to radial geodesics
in the Schwarzschild spacetime.

Our results show that the presence of a quadrupole can drastically affect the motion of test particles in a such a way that it is possible to determine 
the quadrupole by studying the properties of the test particle trajectories. If we were to compare our results with observational data from  astrophysical compact
objects, we would immediately notice that an important astrophysical parameter is missing in our analysis, namely, the rotation. We expect to perform such
an analysis in a future work by using a stationary generalization of the $q-$metric as, for instance, the one derived in \cite{tq14}.

\begin{acknowledgments}
We would like to thank the members of the GTD-group at the UNAM for stimulating discussions and interesting comments.
This work was partially supported by DGAPA-UNAM, Grant No. 113514, and Conacyt, Grant No. 166391. 
We acknowledge the support through a Grant of the Target Program of the MES of the RK. 

\end{acknowledgments}

\appendix*

\section{Calculation of the ADM mass}

Consider the space-time $({\cal M}, g)$  foliated  by  a  family $(\Sigma_t)_{t \in \Re}$ of  
spacelike hypersurfaces. Let 
    \be
     \mathscr{S}_t \equiv \partial \mathscr{V}\; \cap\; \Sigma_t ,
    \ee
where $\partial \mathscr{V}$ is  the boundary  of  a  domain $\mathscr{V}$ ($\partial \mathscr{V}$ 
is  assumed  to  be a timelike hypersurface). The  {\it ADM mass}  of  the  slice $\Sigma_t$ is defined 
 by
    \be
     M_{\text{ADM}} =\frac{1}{16 \pi } \lim_{ \mathscr{S}_t  \rightarrow  \infty }
                     \oint_{\mathscr{S}_t} {[D^j h_{i j} - D_i( f^{k l} h_{k l})] S^i \sqrt{s}\, d^2y}   
                     \label{eq:ADMdef}
 \ee
where  $f$ is  a  flat background   metric on  $\Sigma_t$,  $h$  is the  induced  metric of  the 
hypersurface $\Sigma_t$, $D$ stands  for  the connection  associated with  the  metric $f$,  $S^i$ stands  
for  the components of  unit normal  to $\mathscr{S}_t$,  $ \sqrt{s}\, d^2y$ denotes  the  surface element  
induced  by the  spacetime metric  on $\mathscr{S}_t$ with  $s_{ab}$ being the  induced  metric, $y^a=(y^1,y^2)$ are some  
coordinates on $\mathscr{S}_t$ and $s \equiv \det{s_{ab}}$.  

Let  us  take  for  $\Sigma_t$ the  hypersurface  of  constant coordinate $t$. Then, in the case of the $q-$metric 
(\ref{zv}) we  have    for the induced  metric on  the hypersurface 
$\Sigma_t$
 \bea 
  h_{i j}= \text{diag}\Bigg[
         \left(1-\frac{2m}{r}\right)^{-q-1} \left( 1 + \frac{m^2 \sin^2{\theta}}{r^2-2mr} \right)^{-q(2+q)} &,
        r^2 \left(1-\frac{2m}{r}\right)^{-q} \left( 1 + \frac{m^2 \sin^2{\theta}}{r^2-2mr} \right)^{-q(2+q)},
        \nonumber\\
        & r^2 \sin^2{\theta}\left(1  -\frac{2 m}{r}\right)^{-q}
           \Bigg],
 \eea 
whereas  for  the flat  metric  we  obtain
  \be
    f_{i j}=\text{diag}\Big[1,r^2,  r^2 \sin^2{\theta} \Big].
  \ee 
	
For $\mathscr{S}_t$ we  take the  sphere $r=\text{const}$ on the  hypersurface  $\Sigma_t$. Then 
$y^a=(\theta, \varphi)$, $\sqrt{s} =r^2 \sin{\theta}$ and $S^i= \delta^i_r$. Accordingly,  
Eq.(\ref{eq:ADMdef}) becomes
    \be 
     M_{\text{ADM}} =\frac{1}{16 \pi } \lim_{ r \rightarrow  \infty }
                     \oint_{r=\text{const}} {[D^j h_{r j} - D_r( f^{k l} h_{k l})] r^2 \sin{\theta}
                     d \theta d \varphi} 
                     \label{eq:ADMpart},
 \ee
where 
 \be 
     D_r( f^{k l} h_{k l})= \frac{\partial}{\partial r} (f^{kl}h_{kl}) 
                          = \frac{\partial}{\partial r}\left( h_{rr} + r^{-2}h_{\theta\theta}
                          + r^{-2}\sin^{-2}\theta h_{\varphi\varphi}\right). \label{eq:covder1}
 \ee 
and
   \be 
    D^{j}h_{rj} = f^{jk}D_{k}h_{rj}=\frac{\partial}{\partial r}h_{rr} + \frac{2}{r}h_{rr}
                -\frac{1}{r^3}h_{\theta\theta} -\frac{1}{r^3\sin^2{\theta}}h_{\varphi\varphi}.
                 \label{eq:covder2} 
   \ee 
Here we used the fact that $f^{k l} h_{k l}$ is  a  scalar  field.
In obtainging Eq. (\ref{eq:covder2}), we used the formula
   \bea 
                D_r h_{rr}  &=  \frac{\partial}{\partial r}h_{rr} - 2 \Gamma^r_{rr}h_{rr},\\
    D_{\theta} h_{r\theta}  &= \frac{\partial}{\partial \theta}h_{r\theta} 
                             - \Gamma^{\theta}_{\theta r}h_{\theta\theta},
                             - \Gamma^{r}_{\theta \theta}h_{rr},\\
    D_{\varphi} h_{r\varphi} &= \frac{\partial}{\partial \varphi}h_{r\varphi}
                              - \Gamma^{\varphi}_{\varphi r}h_{\varphi \varphi}
                              - \Gamma^{r}_{\varphi \varphi}h_{r r},
   \eea 
with  the  non-zero  components  of the  Christoffel symbols  of  the  connection $D$ with respect  to the  
coordinates $x^i$ given  by
  \be 
  \Gamma^{r}_{\theta  \theta} = -r,    \quad    
  \Gamma^{r}_{\varphi\varphi} = - r\sin^2{\varphi}, \quad 
  \Gamma^{\theta}_{r\theta}  = \Gamma^{\varphi}_{r \varphi} = \frac{1}{r},\quad 
  \Gamma^{\theta}_{\varphi\varphi}=-\cos{\theta}\sin{\theta}, \quad
  \Gamma^{\varphi}_{\theta\varphi}={\cot{\theta}}.
   \ee 
	
Accordingly, from Eq.(\ref{eq:covder1}) and Eq.(\ref{eq:covder2}), after a simple but tedious calculation,  
one obtains  
that 
   \be 
			D^j h_{r j} - D_r( f^{k l} h_{k l}) = \frac{4m(q+1)}{r^2}
                                          -q(2+q)m^2\left(1 - \frac{2m}{r}\right)^{-q-2} 
                                          \left( \frac{3}{r^3} + \frac{2mq}{r^4}\right) \sin^2{\theta},
                                          \label{eq:integrand}
   \ee 
where  we  have used   the  first-order  approximation 
   \be 
    \left( 1 + \frac{m^2 \sin^2{\theta}}{r^2-2mr} \right)^{-q(2+q)} \approx 
    \left( 1 -\frac{q(2+q) m^2 \sin^2{\theta}}{r^2-2mr} \right).
   \ee 
Thus,  by  substituting Eq.(\ref{eq:integrand}) into Eq.(\ref{eq:ADMpart}) and calculating  the limit after 
integration we  obtain $M_{\text{ADM}}= m(1 + q)$.


\begin{thebibliography}{99}


\bibitem{kerr63} R. P. Kerr, 
Phys. Rev. Lett.  {\bf 11} (1963) 237.

\bibitem{def78} F. de Felice, 
Nature {\bf 273} (1978) 429.

\bibitem{cal79} M. Calvani and L. Nobili, 
Nuovo Cim. B {\bf 51} (1979) 247.

\bibitem{rud98} W. Rudnicki, 
Acta Phys. Pol. {\bf 29} (1998) 981 .

\bibitem{penrose} R. Penrose, Riv. Nuovo Cim. {\bf 1} (1969) 252.

\bibitem{hawking} S. W. Hawking, G. F. R. Ellis, {\it The Large Scale Structure of Space-Time} 
(Cambridge University Press, Cambridge, 1973).

\bibitem{naked} P. S. Joshi, {\it Gravitational Collapse and Spacetime Singularities} 
(Cambridge University Press, Cambridge, 2007).

\bibitem{zip66} D. M. Zipoy, 
J. Math. Phys. {\bf 7} (1966) 1137.

\bibitem{voor70} B. Voorhees, 
Phys. Rev. D {\bf 2} (1970) 2119.

\bibitem{pqr11a}
  D.~Pugliese, H.~Quevedo and R.~Ruffini,
  Phys.\ Rev.\  D {\bf 83}, 024021 (2011).

  \bibitem{pqr11b}
  D.~Pugliese, H.~Quevedo and R.~Ruffini,
  Phys.\ Rev.\ D {\bf 83}, 104052 (2011).


\bibitem{solutions}
H. Stephani, D. Kramer, M. A. H. MacCallum, C. Hoenselaers, and E. Herlt, 
{\it Exact Solutions of Einstein's Field Equations} (Cambridge University Press, Cambridge, 2003).

\bibitem{quev10} H. Quevedo,  Gen. Rel. Grav. {\bf 43} 1141 (2011).


\bibitem{quev90} H. Quevedo, 
Forts. Physik {\bf 38} (1990) 733.

\bibitem{quev11} H. Quevedo, Int. J. Mod. Phys. D {\bf 20} 1779 (2011). 

\bibitem{mala04} D. Malafarina, Conf. Proc. C0405132, 273  (2004).


\bibitem{wald} R. Wald, {\it General Relativity} (University of Chicago Press, Chicago, 1984).

\bibitem{tod11} P. Tod, Gen. Rel. Grav. {\bf 43}, 1855 (2011).
e
\bibitem{ger} R. Geroch, 
{J. Math. Phys.} {\bf 11} (1970) 1955;
{J. Math. Phys.} {\bf 11} (1970) 2580 .

 
\bibitem{par85} S. Parnovsky,
 Zh. Eksp. Teor. Fiz. {\bf 88}, 1921 (1985); JETP {\bf 61}, 1139 (1985). 

\bibitem{zv1} D. Papadopoulos, B. Stewart, L. Witten, Phys. Rev. D {\bf 24}, 320  (1981). 
 
\bibitem{zv2} L. Herrera and J. L. Hernandez-Pastora, J. Math. Phys. {\bf 41}, 7544  (2000).

\bibitem{zv3} L. Herrera, G. Magli and D. Malafarina, Gen. Rel. Grav. {\bf 37}, 1371  (2005).

\bibitem{zv4} N. Dadhich and G. Date, (2000),  arXiv:gr-qc/0012093 


\bibitem{zv5} H. Kodama and W. Hikida, Class. Quantum Grav. {\bf 20}, 5121  (2003).



\bibitem{cpmj12} A. N. Chowdhury, M. Patil, D. Malafarina, and P. S. Joshi, Phys. Rev. D {\bf 85}, 104031 (2012).

\bibitem{qp89} H. Quevedo and L. Parkes, Gen. Rel. Grav. {\bf 21}, 1047 (1989).


 \bibitem{pqr13b}
  D.~Pugliese, H.~Quevedo and R.~Ruffini,
  Phys.\ Rev.\ D {\bf 88},  024042  (2013).

\bibitem{pq15}
  D.~Pugliese and H.~Quevedo,
  Eur. Phys. J. C {\bf 75},  234  (2015).



\bibitem{lq12}
O. Luongo and H. Quevedo, \emph{Toward an invariant definition of repulsive gravity} in Proceedings of the Twelfth Marcel
Grossmann Meeting on General Relativity, edited by T. Damour, R.T. Jantzen, R. Ruffini (World Scientific, Singapore, 2012), Part B, pp. 1029 - 1031; 
arXiv[gr-qc]:1005.4532

\bibitem{lq14}
O. Luongo, H. Quevedo, Phys. Rev. D, {\bf 90}, 8, 084032, (2014).


\bibitem{tq14} S. Toktarbay and H. Quevedo, Grav. \& Cosm. {\bf 20}, 252 (2014).



\end{thebibliography}
\end{document}